\documentclass{aastex62}

\shorttitle{The motion of stars in the Pleiades}
\shortauthors{Danilov \& Seleznev}

\begin{document}

\title{\bf On the Motion of Stars in the Pleiades according to Gaia DR2 Data}

\correspondingauthor{Vladimir M. Danilov}
\email{vladimir.danilov@urfu.ru}

\author{Vladimir M. Danilov}
\affil{Ural Federal University \\
620002, 19 Mira street, \\
Ekaterinburg, Russia}

\author{Anton F. Seleznev}
\affil{Ural Federal University \\
620002, 19 Mira street, \\
Ekaterinburg, Russia}

{
\footnotesize
It is a preprint of an article accepted for publication in Astrophysical Bulletin \copyright 2020 Pleiades Publishing, Ltd.
http://pleiades.online/
}








\begin{abstract}

We used Gaia\,DR2 data on the coordinates, proper motions,
and radial velocities of stars in regions with radius \mbox{$d\degr=2\fdg5$}
and \mbox{size $60\degr\times60\degr$} around the cluster center in order to
estimate several parameters of the Pleiades cluster.
With the data on stars of magnitudes \mbox{$m_G\le 18^{\rm m}$},
we constructed the density maps and profile, luminosity and mass functions
of the cluster, determined the cluster radius,
$10\fdg9\pm0\fdg3$ ($26.3\pm0.7$~pc), and the radius of its core,
$2\fdg62$ ($6.24$~pc), and obtained estimates for the number
of stars in the cluster, $1542\pm121$, and their mass, $855\pm104M_{\odot}$;
numbers of stars in the core of \mbox{the cluster, $1097\pm77$,}
and their \mbox{mass $665\pm71M_{\odot}$.}
Distribution of stars with $m_G<16^{\rm m}$ at distances $r_s$ from
the cluster center in three-dimensional space of $r_s<1$\,pc and
at $r_s$\,$\sim$\,$1.4$--$5$\,pc contains radial density waves.
Based on the data on stars with $m_G<16^{\rm m}$, we determined
the average rotation velocity of the core of the cluster
\mbox{$v_c=0.56\pm0.07$\,km\,s$^{-1}$} at distances $d$ in the sky plane
$d\le4.6$\,pc from its center.
The rotation is ``prograde'', the angle between the projection of the axis
of rotation of the cluster core onto the sky plane and the direction
to the North Pole of the Galaxy is {$\varphi=18\fdg8\pm4\fdg4$,}
the angle between the axis of rotation of the cluster core and the sky plane
is {$\vartheta=43\fdg2\pm4\fdg9$,} the rotation velocity
of the cluster core at a distance of {$d\simeq 5.5$\,pc}
from its center is close to zero: {$v_c=0.1\pm0.3$\,km\,s$^{-1}$.}
According to the data on stars with {$m_G<17^{\rm m}$},
the velocity of the ``retrograde''\, rotation of the cluster at a distance of
{$d\simeq 7.1$\,pc} from its center is {$v_c=0.48\pm0.20$\,km\,s$^{-1}$,}
the angle \mbox{$\varphi=37\fdg8\pm26\fdg4$.}
The dependences of moduli of the tangential and radial components
of the velocity field of the stars of the cluster core in the sky plane
on the distance $d$ to the center of the cluster contain
a number of periodic oscillations.
The dispersions of the velocities of the stars in the cluster core $\sigma_v$
increase on average with an increase in $r_s$, which, like
the radial density waves and the waves of oscillations
of the velocity field in the sky plane, indicates the non-stationarity
of the cluster in the field of regular forces.
The Jeans wavelength in the cluster core decreases,
and the velocity dispersion of the stars in the core under the Jeans
instability increases after taking into account the influence
of the external field of the Galaxy on the cluster.
The region of gravitational instability in the Pleiades cluster
is located in the interval \mbox{$r_s\,$=\,$2.2$--$5.7$\,pc}
and contains {$39.4$--$60.5$\%} of the total number of stars
in the considered samples of cluster stars.
 Estimates of the Pleiades dynamic mass and tidal radius are obtained.
\end{abstract}


\section{INTRODUCTION}
The collection of sufficiently accurate and complete data
on the motions and coordinates of stars in the Galaxy,
recently obtained within the framework of the Gaia project,
makes it possible to verify the theoretical conclusions
formulated some time ago in stellar dynamics \citep{A1,A2}
on the non-stationarity of open star clusters (OSC),
on gravitational instability of cluster cores on the example
of the nearby Pleiades cluster. Among the important tasks
is the study of the structure and internal kinematics of
the Pleiades, since diagnostics of the dynamic state of an OSC
is based on the results of statistical and kinematic studies of
OSCs, on the estimates of the general and local dispersions
of stellar velocities, tidal radii, and total and virial masses of OSCs.

The internal kinematics of OSCs is poorly studied,
which is mainly due to the low accuracy of the previously used data
on proper motions, radial velocities and distances
of {member stars} of OSCs from the Sun. Let us note here several works.
Based on data on proper motions of stars in eight OSCs, \citet{A3}
considered the dependences of the stellar velocity dispersion
on the average mass and radial distance from the cluster center.
In most clusters, such dependences were not found,
isotropy of velocities in the sky plane was observed
in all considered clusters, excluding NGC\,2516, in which the radial
and tangential components of the stellar velocity dispersion
differ in the outer ($r>2$~pc) regions of the cluster.
Typical errors of the root-mean-square deviations
of proper motions of stars from the mean for the considered OSCs
ranged from $0\farcs01$ to $0\farcs05$ for 100~years
(see Table 2 from~\citet{A3}).

The Pleiades and Praesepe clusters were studied in~\citet{A4,A5},
respectively. After identifying the members of the cluster
(with probability $p\ge0.3$), it was found that the mass
of the Pleiades cluster is {$M_{\rm cl}\simeq 800 M_{\odot}$}
(taking into account the contribution of binary stars not resolved
by observations, the value of which was estimated at 15\%
of the "observed" cluster mass $M_{\rm cl}$ obtained after
a constructing the cluster mass function (MF)).
According to~\citet{A4} for the Pleiades, the tidal radius $R_t$ is
13.1\,pc. MF of the cluster turned out
to be flatter in the core than in the halo (see Fig. 10 from~\citet{A4}).
This fact indicates a decrease in the fraction of low-mass stars
with \mbox{$M\le 0.5 M_{\odot}$} due to their dissipation
during stellar approaches, but needs to be confirmed
using more accurate information about the membership of stars
to a cluster at large distances from its center.
The same calculations for the Praesepe cluster performed in~\citet{A5}
lead to the estimates
\mbox{$M_{\rm cl}\simeq 600 M_{\odot}$,} \mbox{$R_t\simeq 12$~pc}
for $p>0.2$ (the contribution of unresolved binary stars was not
taken into account); MF in the stellar mass interval
\mbox{$0.15$--$1.0 M_{\odot}$} is approximately similar to the MF
for the Pleiades (see Fig. 12 from~\citet{A5}), the differences in MF
in the core and halo of the Praesepe cluster also indicate
to reduce the proportion of low-mass stars in the cluster core
compared to the halo (as in the Pleiades cluster).
When considering the stars---the most probable members of the OSC,
the papers~\citet{A4,A5} also noted the ellipticity
of the outer parts of the Pleiades and Praesepe clusters,
which indicates the influence of the tidal field of the Galaxy
at their periphery.

The internal kinematics of the OSC $\alpha$~Perseus was considered
in~\citet{A6} from the data on the proper motions and coordinates
of the {star-members} of the cluster. A conclusion is made
about the general compression of the cluster, an estimate
of the critical value of its tidal density is given
{$0.66 M_{\odot}/$pc$^3$.} About half of the most massive cluster stars
(with spectral types earlier than G) are located inside
a sphere with a radius of 10.3~pc. Arguments in favor of
the gravitational coupling of this OSC are presented.

\citet{A7} used the Hipparcos catalog data on proper motions
and parallaxes known for the cluster members with an accuracy
of at least 5 milliseconds of arc per year and 3 milliseconds of arc,
respectively, to study the internal kinematics of the Hyades cluster.
A correlation was noted between
the tangential velocity component and parallax, which indicates
a possible rotation of the cluster. The axis of rotation
is perpendicular to the direction of the Hyades apex
on the celestial sphere.

\citet{A8} investigated the internal kinematics of the Pleiades,
Praesepe and M\,67 clusters by the data on the coordinates
in the sky plane and the proper motions of the cluster member stars.
Periodic oscillations are identified on the constructed
radial dependences of the values of the moduli of the tangential
and radial projections of the velocities of the stars relative
to the cluster center. An increase in these moduli with distance
to the center of the cluster, as well as the presence
of waves of oscillations of the velocity field in the spectra
of oscillations of the OSCs, considered in~\citet{A8}, indicates
the nonstationarity of these clusters in the field of regular forces.
The parameters of the selected oscillations were used to estimate
the total masses of the Pleiades, Praesepe and M\,67.
For the Pleiades cluster in~\citet{A8}, the following estimates
were obtained by this method: \mbox{$R_t
= 9.5 \pm$ 0.5~pc,} \mbox{$M_{\rm cl} = 330 \pm 55 M_{\odot}$.}

When studying the parameters of the oscillation generation regions
in the OSC models, as well as in the Pleiades, Praesepe and M\,67 clusters,
~\citet{A2} used the condition of gravitational instability
of the cores of star clusters written for an isolated cluster.
It is of interest to refine this condition for the case of OSCs
moving in the field of forces of the Galaxy. It may also be interesting
to discuss the reasons for the formation of ``cool''\, nuclei in the OSCs.

The goals and objectives of this work are to study the structure
of the open star cluster Pleiades in the spaces of coordinates
and velocities of stars, diagnose the dynamic state of this cluster
based on the data of studying its spatial structure
and internal kinematics, as well as estimating the general
and local dispersions of stellar velocities, tidal radii,
total and dynamic masses of the Pleiades cluster for samples
of its members with different limiting magnitudes:
{$m_G<15^{\rm m},16^{\rm m},17^{\rm m}$.}

The results obtained will make it possible to elucidate
the structural features of the Pleiades in coordinate and velocity
spaces, check a number of theoretical conclusions about the OSCs
made in stellar dynamics, and outline further directions of research
on the structure, internal kinematics, and dynamics of the OSCs.

\section{BASIC FORMULAS AND CALCULATION TECHNIQUES}
\label{R2}

\citet{A9} obtained his formula (7.2) for the Jeans wavelength $\lambda_J$
of an isolated homogeneous spherical cluster of stars
from the condition that the time $t_{\rm coll}$ of the compression of
a cold spherical region of radius $\lambda$ under the action of gravitational
forces is equal to the time {$t_{\rm esc}=\lambda/\sigma_v$}
of the exit of a star from such a region with the velocity dispersion
$\sigma_v^2$.
This formula was used in~\citet{A2} to estimate the Jeans mass $M_J$
in the cluster core related to the mass of its core
$M_c$ {($M_J=qM_c, q\le 1, q=$const),} and to calculate $\sigma_v^2$
at a distance of $r_c$ from the center of the cluster.
According to~\citet{A2}:
\begin{equation}
\label{eq1} \sigma_{v}^2 = 32Gq^{2/3}(\rho_c M_c^2)^{1/3}/(3\pi),
\end{equation}
where $r_c$ was taken to be equal to the distance $r_s$
from the center of the cluster, at which the modulus of the gradient
of the spatial density of the number of stars (and mass density)
of the cluster sharply decreases when passing along $r_s$
from the core to the halo or to an intermediate zone
of increased density in the cluster, $G$ is the gravitational constant,
$\rho_c$ is the mass density at a distance $r_c$ from
the cluster center, $q$ values for six OSC models are given
in Table~\ref{Tab 1} of~\citet{A2}, $M_c$ is the mass
of the cluster core, taken equal to the sum of the masses of stars
with distances {$r_s\le r_c$} from the center of the cluster.

Let $\omega_h$ be the frequency of small homological oscillations
of the model of a spherical homogeneous cluster of stars moving
in the force field of the Galaxy in a circular orbit
with an angular velocity $\omega$ relative to the galactic center.
According to~\citet{A10}, $$\omega_h=\sqrt{2q_0^2/3+K}.$$
Here~ {$q_0^2~=~\alpha_1~+~\alpha_3~+~3\beta$, where~ $\alpha_1$,
$\alpha_3$} are the constants characterizing the force field
of the Galaxy~~ in the~~ vicinity~~ of the circular~~ orbit~~
of the cluster~~\citep{A11}, {$\beta~=~GM_{\rm cl}/R_{x}^3$,}
the value~~ {$K~~=~~2\left(4\omega^2\,+\,q_0^2\,-\,9p^2/(2I_0^{3/2})\right)/\,3$,}\\
\mbox{$p^2=GM_{\rm cl}\,(0.6\,M_{\rm cl})^{3/2}$; $M_{\rm cl}$}
is the mass of the cluster, {$R_{x}$} and \mbox{$I_0=0.6M_{\rm
cl}R_{x}^2$} are the radius and moment of inertia of the considered
cluster model, respectively. The quantities $\omega$, $\alpha_1$, $\alpha_3$ are
determined in our work using the model of the Galaxy potential~\citep{A12}.

Let {$\gamma=t_{\rm coll}/t'_{\rm coll}$,} where $t'_{\rm
coll}$ is the compression time in the vicinity of the cluster center
of the cold region with radius $\lambda$ in the total field of forces
of stars in this region and the Galaxy.
Note that {$\gamma=\omega_{h}/\omega_{h,0}$,} where
$\omega_{h,0}$ is the frequency of small homologous oscillations
of an isolated cluster {($\omega_h=\omega_{h,0}$} at
\mbox{$\omega=\alpha_1=\alpha_3=0$).}
The estimate of the parameters of the Jeans instability
is closely related to the estimates of the frequency
of the resulting oscillations of the gravitating system,
which are formed as a result of the development of such instability.
Let $\tau_{\rm v.r.}$ be the initial time of violent relaxation of the cluster.
The value $\tau_{\rm v.r.}$ Is obtained according to the formula from~\citep{M6}:
{$\tau_{\rm v.r.}\simeq 2.6 \overline{t}_{\rm cr}$}, where
$\overline{t}_{\rm cr}$ is the average initial time of a star crossing
the cluster. Small perturbations of the phase density in the cluster core
under the influence of such instability increase rather quickly,
during the time {$\Delta t\simeq0.5\tau_{\rm v.r.}$}
(see, for example, the second column of Table 2 and the figure
from~\citet{A13} for the time dependences of the relative perturbations
\mbox{$\Psi_i(t)$} of the coarse-grained phase density in OSC models 2, 4, 6
at \mbox{$i=1$}), capturing more and more parts of the cluster and
passing into stable oscillations of the entire system as a whole
(see the transition of the dependencies $\Psi_i(t)$ to the ``plateau''\, at
{$t>0.5\tau_{\rm v.r.}$} in OSC models 2, 4, 6 from~\citet{A13},
Fig.~5 for the dependences $\Psi_i(t)$ in OSC models 2, 4 at
$i=3,4,5$ from \cite{A14}, as well as theoretical estimates
of the frequencies of stable homological and nonhomological oscillations
of OSC models in \cite{A10,A1}).

From the condition {$t'_{\rm coll}=t_{\rm esc}$} we find:
{$\lambda_J=t_{\rm coll}\sigma_v/\gamma$} and
{$\sigma_v=\gamma\lambda_J/t_{\rm coll}$.} Because {$\gamma>1$,}
then after taking into account the action of the Galactic field on the cluster,
$\lambda_J$ decreases, and $\sigma_v$ increases.
Taking into account the condition {$t'_{\rm coll}=t_{\rm esc}$}
and repeating the calculations performed in~\citet{A2}, for a non-isolated
cluster we find:
\begin{equation} \label{eq2} \sigma_{v}^2 =
32\gamma^2Gq^{2/3}(\rho_c M_c^2)^{1/3}/(3\pi).\end{equation}
When using the data on the considered samples of the Pleiades cluster stars
up to different limiting magnitudes $m_G$, the value {$\gamma\simeq
1.080$--$1.163$.} The value $\gamma$ increases with the size
of the cluster's gravitational instability region.
According to formula (11) from~\citet{A10}, the following equation
can be written to estimate the dynamic dispersion of stellar velocities
in the OSC from the data on the structural parameters of the cluster:
\begin{equation} \label{eq3}
 \sigma^2_{v,d} \simeq
\frac{1}{2}\left\{\frac{1}{M_{\rm cl}} \left[-W + \frac{1}{3}
 \left(\alpha_1 + \alpha_3
 \right) I\right] + \frac{U(0)}{2}\right\},
\end{equation}
where $W$, $I$ and $U(0)$ are the potential energy of the cluster,
its moment of inertia and the potential at the center of the cluster,
respectively. Estimating the value of $\sigma^2_{v,d}$ takes
into account the influence of the force fields of the cluster
and the Galaxy, as well as the nonstationarity of the cluster
on the velocity dispersion of stars in the cluster.
Assuming the value $\sigma^2_{v,d}$ equal to the same value
obtained for the cluster model in the form of a homogeneous ball
with cluster mass, we find the equation for calculating the radius
$R_x$ of this ball:
\begin{equation} \label{eq4}
 R_x^3-\frac{10\sigma^2_{v,d}R_x}{\alpha_1+\alpha_3}+
 \frac{27GM_{\rm cl}}{4(\alpha_1+\alpha_3)}=0.
\end{equation}
This equation has three real roots, which are easily found by
Cardano's formulas~\citep{A15}. The smallest positive root
is used to calculate the frequencies {$\omega_h$, $\omega_{h,0}$}
and the value $\gamma$ (see above). When using the data
on the considered samples of the Pleiades cluster stars
up to different limiting magnitudes $m_G$, the values $R_x$ and $W/W_x$
take the following values: \mbox{$R_x\simeq 3.487$--$4.328$}~pc
and \mbox{$W/W_{x}\simeq 1.068$--$1.018$,} respectively,
where $W_{x}$ is the potential energy of a homogeneous ball
with radius $R_x$ and mass $M_{\rm cl}$.

To estimate the dynamic mass $M_d$ of a non-stationary and non-isolated
Pleiades cluster, we used the formula (2) from~\citet{A2},
received earlier~\citep{A18} based on data from numerical experiments
and in the work of~\citet{A8} converted to the more computationally
convenient form:
\begin{equation}
\label{eq5}
M_{d}=\frac{2\overline{R}R_u}{G(\overline{R}+R_u)}\left[2\sigma^2_v-(\alpha_1+\alpha_3)\overline{r_s^2}/3\right],
\end{equation}
where $\sigma^2_v$ is the velocity dispersion of stars in the cluster,
$\overline{R}$ is the average radius of the cluster (the average distance
between two stars in the cluster; averaging was performed over all pairs
of stars in the cluster), {$R_u=\langle1/r_s\rangle^{-1}$,
$r_s$} is the distance of a star from the cluster center of mass
(angle brackets mean averaging over all cluster stars),
$\overline{r_s^2}$ is the mean square of the star's distance
from the center of the cluster.
The value $\sigma^2_v$ used in (\ref{eq5}) can be obtained
both from observational data on the peculiar motions of stars,
and using the relation \mbox{$\sigma^2_v=\sigma^2_{v,d}$}
(taking into account the results of numerical modeling
of the dynamics of the OSC). In the case of
{$\sigma^2_v=\sigma^2_{v,d}$} formulas (\ref{eq3}) and (\ref{eq5})
allow solving the problem of determining the total mass of the OSC
without using data on the velocities of the stars---members
of the cluster.

To construct the radial dependences of the moduli of the tangential
and radial projections of the velocities of the stars $V_{t}$
and $V_{d}$, obtained from the data on the proper motions
of these stars relative to the center of the Pleiades cluster
in the sky plane, we used the formulas and methods
from~\citet{A8}.

The radial dependences of the quantities $V_{t}$, $V_{d}$
and the apparent density of the number of stars $F(d)$ in the cluster,
as in \cite{A8}, were obtained from the stars closest
to the circle of radius ${d}$ in the sky plane
centered at the center of the cluster. To calculate the values
$V_{t}$, $V_{d}$, $F(d)$  for {$d$ = const} with a step
{$\Delta \varphi=1\degr$} along the angle $\varphi$ relative
to the center of the cluster in the sky plane,
the coordinates of the nodal points were determined.
Then, using six stars closest to each nodal point
(for {$n_{\rm st}=6$}), we determined the values $V_{t}$,
$V_{d}$ and $F(d)$. Further, these values were calculated
as averages over all nodal points (the errors of these averages
are equal to the errors of the values $V_{t}$, $V_{d}$ and $F(d)$).
As in~\citet{A8}, the distributions $F(d)$ were smoothed
by the local weighted regression method~\citep{M16} until
they were reduced to the form of functions monotonically decreasing
in ${d}$, the values of which were taken to be zero for {$d=R_m$,}
where $R_m$ is the radius of the region in which the velocity field
of the cluster stars is studied. To pass from the distribution
$F(d)$ to the spatial density distribution $f(r_s)$, we used
the assumption of the spherical symmetry of the distribution
of stars (and mass) in the cluster, as well as the solution (8.7)
of the Abel integral equation for the function $f(r_s)$
in the book of~\citet{M17}.

The radial dependences on $r_s$ of the dispersions of
the three-dimensional velocities of the stars $\sigma^2_v$
at the distances \mbox{$r_s'\in [0,r_s]$} from the cluster center
were determined from the data on the equatorial coordinates,
parallaxes, and proper motions of stars relative to the center
of the Pleiades cluster. The $\sigma^2_v$ values were calculated
by the formula {$\sigma^2_v=1.5\sigma_{2,v}^2$} under
the assumption of spherical symmetry of the stellar velocity
distribution in the cluster, where $\sigma_{2,v}^2$ is the dispersion
of two-dimensional velocities of motion of stars in the sky plane.

To determine the structural parameters of the cluster and their errors
in Section~\ref{R6}, we used the method described in the works of \cite{A8,A18},
according to which the spatial positions of $6 N_{\rm cl}$ stars
were set using a random number generator in the spherical coordinate
system $(r_s,\theta,\phi)$ of the cluster stars.
The function $f(r_s)$ was used to calculate the probability
distribution density $p_a(r_s)$ of a star hitting the interval
{$r_s\in(0, R_m)$.}
Discrete random variable $r_s$ with a given density $p_a(r_s)$
was distributed in the interval {$r_s\in (0, R_m)$} according
to the method from \citetext{\citealp*[p.~26]{A19}},
and the values $\theta$ and $\phi$---in the intervals
{$\theta \in (0,\pi)$} and {$\phi \in (0, 2\pi)$} with densities
that ensure a uniform distribution of stars by the angles
$\theta$ and $\phi$ for each fixed value {$r_s=r_{s, i}$.}
As a result, sets of values
$({r_{s,i}}, \theta_i, \phi_i)$, {$i=1,..., 6 N_{\rm cl}$.}
were obtained. Each of the six sets of coordinates of $N_{\rm cl}$
stars imitate this cluster of stars. When evaluating
the structural parameters of the OSC, we calculated the average
values of these parameters and the standard deviations
from the average over six sets of coordinates of {$N_{\rm cl}$ stars.}

\begin{figure*}
   \centering
 \includegraphics[scale=0.65]{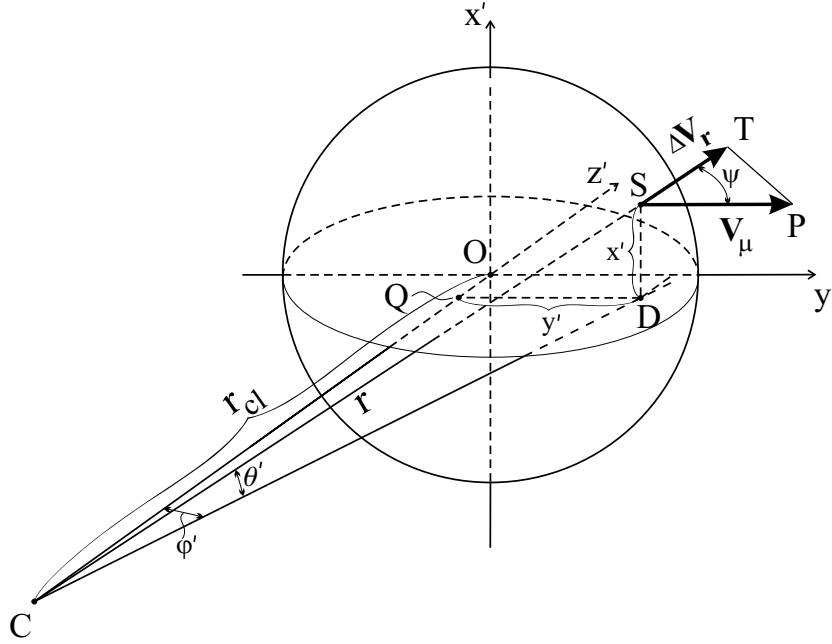}
 \caption{Estimation of the radial velocity increments of OSC stars due to the motion of the cluster perpendicular to the line of sight.}
 \label{Fig1}
\end{figure*}
The Pleiades are a fairly close cluster. Distance to the cluster is
{$r_{\rm cl}\simeq 136.4\pm 0.2$}\,pc, modulus of the cluster
velocity in the sky plane is
{$V_{\mu}\simeq 32.08\pm 0.08$\,km\,s$^{-1}$}
(according to average parallax and average proper motion
of cluster stars according to Gaia data).
In this case, the component of the motion of the cluster,
perpendicular to the line of sight, distorts the radial velocities
of the stars in the leading and lagging parts of the cluster (respectively to this motion), forming,
respectively, positive and negative increments of the radial velocities
of stars from these regions of the cluster. These increments create
the apparent effect of rotation of the cluster about an axis
perpendicular to the direction of motion of the cluster.
Similar distortions can sometimes be seen in the proper motions
of stars due to the motion of the cluster along the line of sight
(the formation of the radiant and apex in the proper motions of the stars).
Let ${\bf V}_{\mu}$ be the projection of the OSC velocity vector
onto the sky plane. To estimate the increments
of the radial velocities of stars due to the motion of the OSC
perpendicular to the line of sight, we consider the right-hand
Cartesian coordinate system $x'$, $y'$, $z'$, the origin of the axes
of which (point $O$) is at the center of the cluster, the $y'$ axis
is directed along the vector ${\bf V}_{\mu}$, the $x'$ axis
is perpendicular to the ${\bf V}_{\mu}$ vector, and the $z'$ axis
is directed from the observer located at the point $C$; the plane
$(x',y')$ coincides with the sky plane.
In Fig.~\ref{Fig1} the circle with the center at the point $O$
denotes the line of intersection of the spherical surface
enveloping the cluster with the plane $(x',y')$, the triangle
$CDS$ is right-angled with a right angle at the vertex $D$.
Directional cosines of the vector ${\bf V}_{\mu}$ relative to
the axes $x'$, $y'$, $z'$ are equal to:
{$\cos \alpha_{x'}=0$,} {$\cos \alpha_{y'}=1$,} \mbox{$\cos
\alpha_{z'}=0$} respectively, see Fig.~\ref{Fig1}.
Let {$r=CS$} be the distance to the star $S$, $r_{\rm cl}=CO$.
The directional cosines of the vector $\Delta {\bf V}_r$ satisfy
the following relations: {$|\cos\beta_{x'}|\le 1$,}
\mbox{$\cos\beta_{y'}=y'/r$,} \mbox{$|\cos\beta_{z'}|\le 1$,} see
Fig.~1, as well as formulas~\mbox{(3.1--8)} from \cite{A15}.
According to~\mbox{(3.1--11)} from \citet{A15}, {$\cos\psi
=\cos\alpha_{y'}$ \mbox{$ \cos \beta_{y'}=y'/r$,} where $\psi$ is
the angle between the vectors ${\bf V}_{\mu}$ and \mbox{$\Delta {\bf V}_r$}
in a right-angled triangle $SPT$ with a right angle at the vertex $T$, and
\mbox{$\Delta V_r=V_{\mu}\cos\psi$,} where $\Delta {V}_r=|\Delta
{\bf V}_r|$, \mbox{$V_{\mu}=|{\bf V}_{\mu}|$~(Fig.~\ref{Fig1}).}
Therefore, \mbox{$\Delta {V}_r=V_{\mu}y'/r$}.
For \mbox{$y'=7$--$10$}\,pc for the Pleiades cluster
$\Delta {V}_r\simeq 1.64$--$2.35$\,km\,s$^{-1}$.
When the sign of $y'$ changes, the sign of $\Delta {V}_r$ also changes.
Therefore, at the opposite (in $y'$) edges of the cluster,
the difference in the radial velocities of the stars due
to the effect considered here reaches values
{$2\Delta {V}_r\simeq 3.3$--$4.7$~km\,s$^{-1}$.}

Data on the rotation of the outer regions of the Pleiades cluster
can be obtained using the radial velocities of the cluster's member stars
located on opposite edges of the cluster near its equator.
The rotation of the cluster core can be studied using
the average proper motions of stars located closer and further
from the cluster center relative to the observer, as well as stars
located in the sky plane to the left and right
(for $l<l_c$ and $l>l_c$), as well as above and below its center
(for $b>b_c$ and $b<b_c$), for example, in the galactic coordinate
system $l$ and $b$, where $l_c$, $b_c$ are the coordinates
of the cluster center. Based on the data on the radial velocities
of stars, such an approach gives the possibility to estimate
the angular velocities $\omega_x$, $\omega_y$, $\omega_z$
of the rotation of the cluster core along the coordinate axes
$x$, $y$, $z$ (the largest contribution to the average proper motions
of such groups of stars are contributed by the stars of the core,
since the apparent and spatial stellar density in the core
is much higher than in the halo and corona of the cluster).

\begin{figure*}
   \centering
 \includegraphics[scale=0.65]{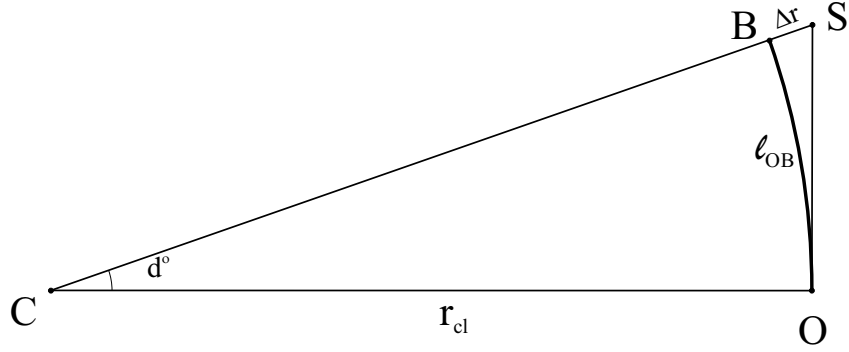}
 \caption{Estimating the increments $\Delta r$ and the arc length $l_{OB}$ near the OSC sky plane.}
 \label{Fig2}
\end{figure*}
Let the axes $x$ and $y$ of a rectangular coordinate system
centered at the center of the cluster are directed towards
increasing coordinates $l$ and $b$, respectively, the $z$ axis
is directed away from the observer, $(x,y)$ is the sky plane.
Because the vector of the angular velocity of rotation of the cluster
{$\Omega =(\omega_x, \omega_y, \omega_z)$}
is directed along the axis of rotation of the cluster, and the equation
of the sky plane is {$z=0$,} then the angle $\vartheta$ between
the vector $\Omega$ and the sky plane, according to
{(3.4--6)} from~\citet{A15}, can be determined by the formula
\begin{equation} \label{eq6}
 \sin \vartheta= \frac{\omega_z}{\sqrt{\omega_x^2+\omega_y^2+\omega_z^2}},
\end{equation}
where $\omega_x$, $\omega_y$, $\omega_z$ are the projections
of the vector $\Omega$ on the coordinate axes $x$, $y$, $z$ respectively.
If the average radii $r_x$, $r_y$, $r_z$ of the circular motion
of groups of stars about the axes $x$, $y$, $z$ are equal to each other,
then the values $\omega_x$, $\omega_y$, $\omega_z$ for these groups
can be replaced by average velocities $\overline{v_{y,z}}$,
$\overline{v_{x,z}}$, $\overline{v_{x,y}}$ of the
circular motion of groups of stars relative to the axes $x$, $y$, $z$,
respectively (since the multiplication of the numerator and denominator
on the right side of expression~(\ref{eq6}) by the radius
of the circle along which the stars move relative to the
coordinate axes does not change the value of~$\vartheta$).
If the values $r_x$, $r_y$, $r_z$ are not equal, then
in formula (\ref{eq6}) it is necessary to use the velocities
of the stars in the samples, leading to approximately the same
values of $r_x$, $r_y$, $r_z$.

The value of $\overline{v_{x,y}}$  can be estimated from the data
on proper motions (and the corresponding velocities $v_{x}$,
$v_{y}$ of motion in the sky plane) of cluster stars
with distances $r<r_{\rm cl}$ and \mbox{$r>r_{\rm cl}$}:
$\overline{{\bf v}_{x,y}}=(\Delta\overline{{v}}_x,$
$\Delta{\overline{v}}_y)$, $\overline{{
v}_{x,y}}=\pm\sqrt{\Delta\overline{{v}}_x^2+\Delta{\overline{v}}_y^2}$,
where $\Delta{\overline{v}}_x=0.5({\overline{v}}_x(r<r_{\rm
cl})-{\overline{v}}_x(r>r_{\rm cl}))$, $\Delta{\overline{v}}_y$
is defined similarly after replacing the index $x$ with $y$
in the expression for $\Delta{\overline{v}}_x$.
The overline in $\Delta{\overline{v}}_x$ and {$\Delta{\overline{v}}_y$
denotes averaging over stars with {$r<r_{\rm cl}$} and {$r>r_{\rm cl}$.}
Estimates of the values of $\overline{v_{x,z}}$, $\overline{v_{y,z}}$
are possible using proper motions, radial velocities and
distances of cluster stars: $\overline{{\bf
v}_{x,z}}=(\Delta\overline{{v}}_x,\Delta{\overline{v}}_z(x))$,
$\overline{{
v}_{x,z}}=\pm\sqrt{\Delta\overline{{v}}_x^2+\Delta{\overline{v}}_z(x)^2}$,
where
$\Delta{\overline{v}}_z(x)=0.5({\overline{v}}_z(x<x_c)-{\overline{v}}_z(x>x_c))$,
$\overline{{\bf
v}_{y,z}}=(\Delta\overline{{v}}_y,\Delta{\overline{v}}_z(y))$,
$\overline{{v}_{y,z}}=\\\pm\sqrt{\Delta\overline{{v}}_y^2+\Delta{\overline{v}}_z(y)^2}$,
where
$\Delta{\overline{v}}_z(y)=0.5({\overline{v}}_z(y<y_c)-{\overline{v}}_z(y>y_c))$.
The overline in $\Delta{\overline{v}}_z$ denotes averaging
over stars with coordinates $x$ and $y$, satisfying the constraints
specified in the corresponding formulas for $\Delta{\overline{v}}_z$.
The values $x_c=0$, $y_c=0$, $z_c=0$ are the coordinates
of the cluster center in the $(x,y,z)$ coordinate system.

In Fig.~\ref{Fig2} the center of the cluster and the observer
are located at the points $O$ and $C$, respectively;
$l_{\rm OB}$ is the length of the arc ${OB}$ of a circle centered
at point $C$ and radius $r_{\rm cl}$, the triangle $COS$
is right-angled with a right angle at the vertex $O$,
the line $OS$ lies in the sky plane.
Let us estimate the value
\mbox{$\Delta L=OS-l_{\rm OB}$.} According to Fig.~\ref{Fig2},
\mbox{$OS=r_{\rm cl}\tan d_{\rm rad}$,} \mbox{$l_{\rm OB}=r_{\rm
cl}d_{\rm rad}$,}
where \mbox{$d_{\rm rad}=d\degr\pi/180\degr$} is
the angle $d\degr$ in radians. Therefore,  $\Delta
L=r_{\rm cl}(\tan d_{\rm rad}-d_{\rm rad})$, $\Delta r=CS-r_{\rm
cl}=r_{\rm cl}(1-\cos d_{\rm rad})/\cos d_{\rm rad}$.

Let $r_{\rm cl}=136$\,pc.  Then with
\mbox{$d\degr=10\degr$--$30\degr$} the value $\Delta
L=0.24$--$7.3$\,pc, $\Delta r = 2.1$--$21.0$\,pc,
$\Delta r/OS=\\0.087$--$0.268$, and with $d\degr=10\fdg9$,
we find that \mbox{$\Delta L=0.32$\,pc,} \mbox{$\Delta r =
2.5$\,pc,} \mbox{$\Delta r/OS=0.095$.}
Thus, for the Pleiades cluster, the values $\Delta L$ and
$\Delta r/OS$ are sufficiently small for the values of
$d\degr\in[0\degr,10\fdg9]$  considered in our work and
the use of the sky plane approximation for estimation
of the structural and dynamic parameters of the cluster
seems quite acceptable.

\section{STAR COUNTING IN THE PLEIADES}
\label{R3} Two different approaches
are possible when studying star clusters on the base
of the Gaia space mission data~\citep{S2,S1}. The first is to select stars that have
a high accuracy in determining parallaxes and proper motions.
In this case, the calculated data quality criteria (filters)
published in the Gaia\,DR2 catalog are used. This approach
makes it possible to study the structural features of clusters
(for example, tidal tails), the internal kinematics and dynamics
of clusters, and determine such characteristics of clusters
as age, distance, and color excess. Unfortunately, this approach
loses a significant part of the cluster members, which
have large errors in determining the parameters. This approach
will always give an undersampling of the cluster members.

The second approach consists of statistical study of the clusters.
It uses the Gaia\,DR2 catalog as a complete survey of the whole sky
(it can be considered complete up to the magnitude
\mbox{$m_G=18^{\rm m}$).} In this case, careful selection
of the likely cluster members is not required. The only purpose
of restrictions on the parameters of stars with this approach
is to reduce the sample size and to reduce the fluctuations
introduced by the field stars into the statistically
determined distribution functions. The necessary condition is that
the imposed constraints should not discard members of the cluster.
The obtained samples, containing field stars and almost
all members of the cluster, are processed by statistical methods;
as a result, we have a density profile, surface density maps,
a luminosity function, and a mass function. Both of these approaches
were used in the study of the open cluster Ruprecht 147~\citep{S11}.

To perform star counts in the Pleiades cluster within
the framework of the second approach, from
the Gaia\,DR2 catalog~\citep{S2,S1},
we selected data on 47195 stars, the parameters of which
satisfied the following restrictions. The coordinates of the stars
are: right ascension $\alpha\in[23\fdg845,89\fdg645]$ and
declination $\delta\in[-5\fdg883,54\fdg117]$. Trigonometric parallaxes
$\pi_{r}\in[4,15]$ milliarcseconds (hereinafter mas),
proper motion in right ascension $\mu_{\alpha}\in[10,30]$ milliseconds
of arc per year (hereinafter mas\,year$^{-1}$), proper motion
in declination  $\mu_{\delta}\in[-55.5,-35.5]$ mas~year$^{-1}$.
No Gaia quality filters were used.

Selection by coordinates and parallaxes gave us a region of space,
which is a fragment of a spherical layer with dimensions
of approximately $160\times160\times160$\,parsec, in the center
of which the Pleiades cluster is located. The selection
by proper motions corresponds to the scatter of stellar velocities
in the sky plane approximately $\pm$7\,km/s relative
to the average velocity of the cluster. It can be concluded
that our sample includes all the stars of the Pleiades cluster
up to the limiting magnitude $m_G=18^{\rm m}$, with the possible
exception of distant parts of its tidal tails and stars
with very erroneous parallax and proper motion.

How strongly can the parameters of stars differ from the average
values for a cluster due to errors of Gaia\,DR2? This can be estimated,
for example, by the brightest stars in the Pleiades. According
to parallaxes, the difference reaches approximately 2.2\,mas,
according to proper motions---about 5.3\,mas\,year$^{-1}$.
The half-width of the intervals for parallaxes and proper motions
used by us to obtain the sample is much larger than these values.
It can be concluded that our sample contains all the Pleiades stars
up to $m_G=18^{\rm m}$ even taking into account possible
crude errors in the parameters of the stars.

The number of stars in the Pleiades decreases sharply at
$m_G>18^{\rm m}$ (this corresponds approximately to the mass
of the star $m\simeq0.16M_{\odot}$). A few cluster stars are visible
up to magnitudes of $m_G=20^{\rm m}$, fainter stars are practically
absent. This is clearly seen in the diagrams ``$\pi_{r}$--$m_G$'',
``$\mu_{\alpha}$--$m_G$'' and ``$\mu_{\delta}$--$m_G$''.
\begin{figure*}
   \centering
 \includegraphics[scale=0.45]{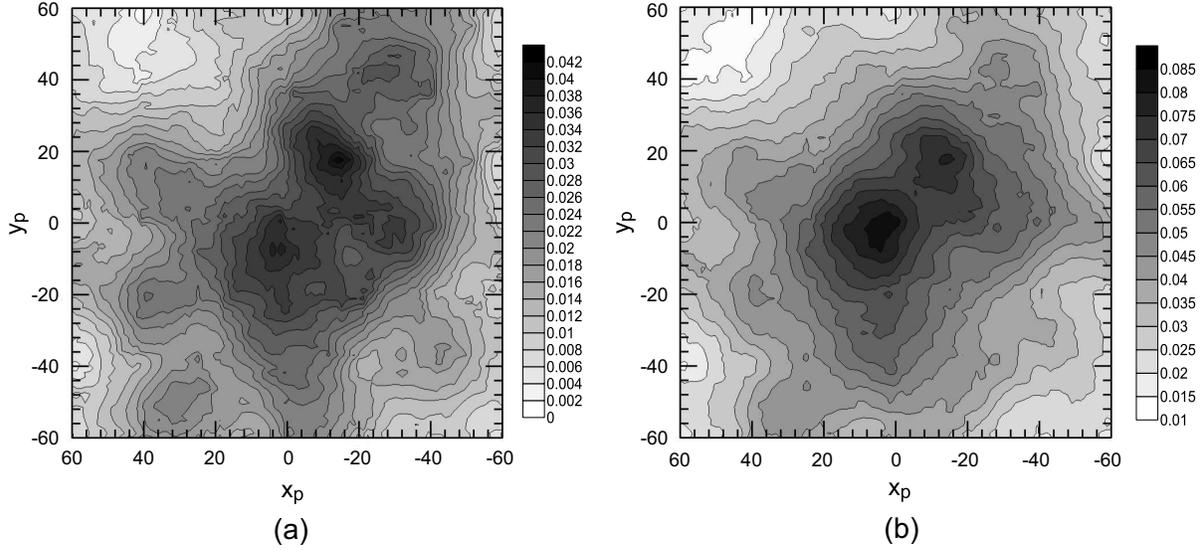}
 \caption{Surface density maps for the central part of the Pleiades cluster
 $(2\degr\times2\degr)$,  $h=20\arcmin$:
 (a) for stars with $m_G\leqslant15^{\rm m}$,
 (b) for stars with $m_G\leqslant18^{\rm m}$.
 The coordinates $x_{p}$ and $y_{p}$ are in arc minutes.}
\label{Fig3}
\end{figure*}
\begin{figure*}
   \centering
 \includegraphics[scale=0.45]{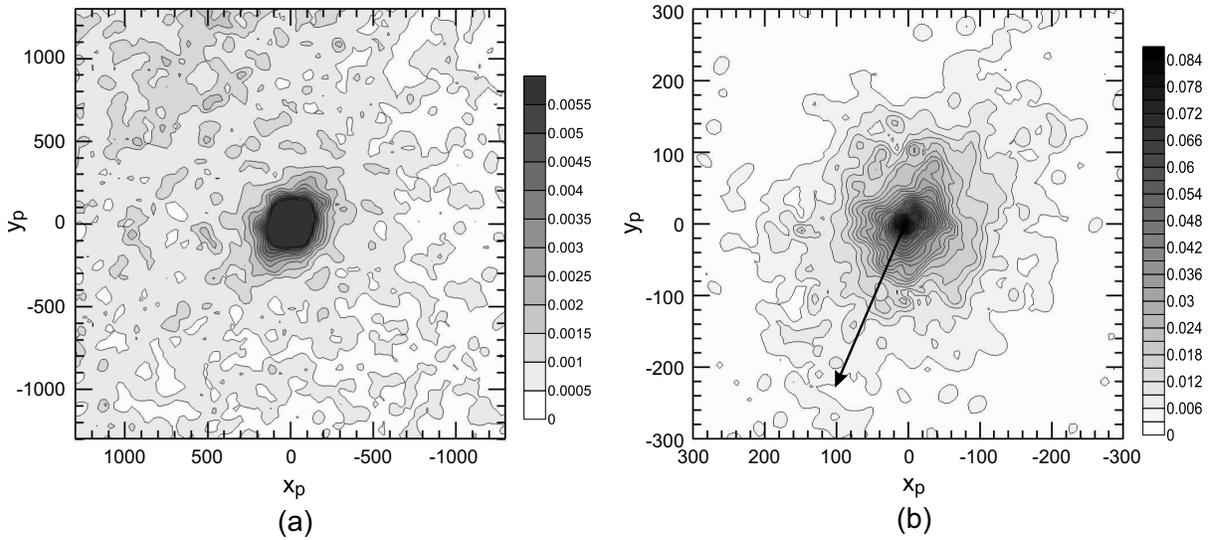}
 \caption{Surface density maps for the vicinity of the Pleiades cluster:
 (a) the region $43\fdg3\times43\fdg3$,
 $h=60\arcmin$; (b) the region $10\degr\times10\degr,$ $h=20\arcmin$;
 the arrow shows the direction of motion of the cluster.
 The coordinates $x_{p}$ and $y_{p}$ are in arc minutes. }
 \label{Fig4}
\end{figure*}

First, we specified the position of the cluster center for
the resulting sample. In this case, the ``kernel density estimator''\, (KDE)
method was used in a one-dimensional version, applied separately
to the equatorial coordinates of the stars in the sample.
In fact, this method is similar to the classical Plummer method
(counting stars in parallel stripes).

With kernel halfwidth $h=0\fdg5$ for stars with
$m_G\leqslant18^{\rm m}$ the coordinates
$\alpha_c=56\fdg69$ and $\delta_c=24\fdg17$ were obtained
for the cluster center, which is different from the values from the database
WEBDA\footnote{\url{https://webda.physics.muni.cz/}}
($\alpha_c=56\fdg75$ and $\delta_c=24\fdg117$).
Note that the position of the center depends
on the limiting magnitude of the stars and can vary
by tens of arc minutes (this is clearly seen in Fig.~\ref{Fig3},
which shows the surface density maps of the central part
of the cluster core for stars of various limiting brightness).
In our case, it makes sense to refine the center to build
a radial density profile (to avoid a decrease in density
in the center of the cluster).

The obtained coordinates of the cluster center were used
to pass to the tangential coordinate system $x_{p}$ and $y_{p}$,
having a pole at the \mbox{point $(\alpha_c,\delta_c)$} \cite[Chapter 20,
Section 4.2]{S12}. Tangential coordinates are used to obtain
distribution functions of stars, from density profiles to
luminosity functions~\citep{S3,S4,S5,S6,S7}.

\begin{figure*}
   \centering
 \includegraphics[scale=0.4]{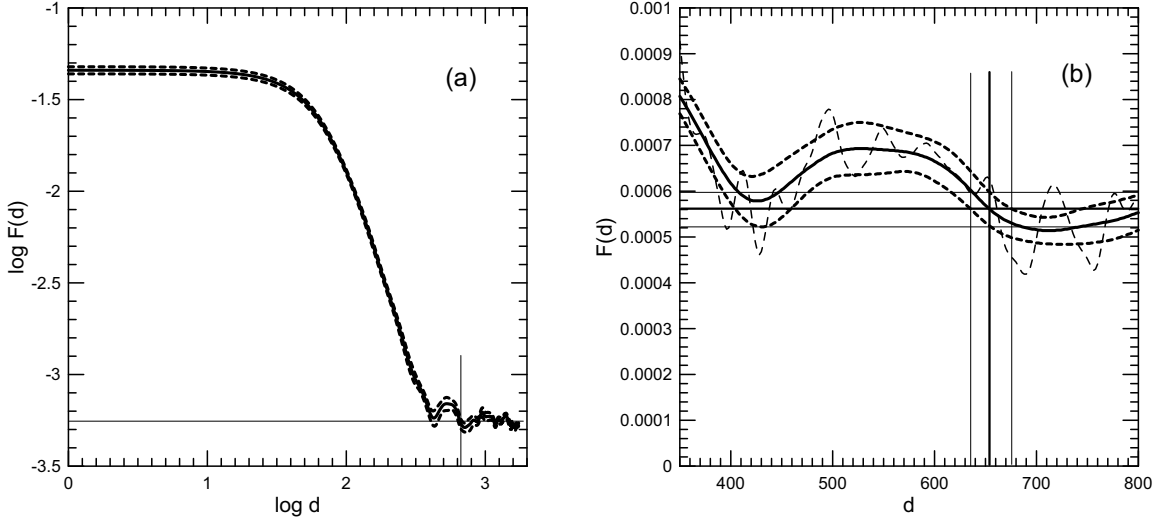}
 \caption{ Surface density profiles of the Pleiades cluster for stars with
 $m_G\leqslant18^{\rm m}$, $h=80\arcmin$. (a) Density profile
 in logarithmic axes (decimal logarithm) is the solid curve line;
 the dashed curves show the $\pm1\sigma$ confidence interval;
 the solid straight lines show the accepted values
 of the cluster radius and average density of background stars;
 (b) density profile in the linear axes is the thick solid curved line,
 the area near the cluster boundary is shown; the thick dashed lines
 show the $\pm1\sigma$ confidence interval; the thin dashed line shows
 the density profile plotted at $h=20\arcmin$; the thick straight
 lines show the assumed values of the cluster radius and average
 density of background stars; thin straight lines illustrate
 the determination of the error of the cluster
 radius and average density of background stars.}
 \label{Fig5}
\end{figure*}

Fig.~\ref{Fig3} shows the surface density maps for
the central part of the cluster core, the region $2\degr\times2\degr$
($h=20\arcmin$). Fig.~\ref{Fig3}a corresponds to the distribution
of stars with magnitude $m_G\leqslant15^{\rm m}$,
Fig.~\ref{Fig3}b also includes fainter stars up to magnitude
$m_G\leqslant18^{\rm m}$. It can be seen that when passing
to brighter stars, the center of the cluster changes its position
by about $20'$ in $x$ and $y$. The density inhomogeneities
in these maps are solely due to the cluster stars,
as the lowest density gradation in these maps is 0.002 stars
per square arc minute, and the average background star density
for our sample is 0.000562 stars per square arc minute (see below).
Note the complex irregular structure of the cluster core
(Fig.~\ref{Fig3}), which is more complicated for the subsystem
of brighter stars with {$m_G\le 15^{\rm m}$,} which indicates
large deviations of the cluster core from equilibrium
in regular field.

Fig.~\ref{Fig4} shows surface density maps for the vicinity
of the Pleiades cluster. Fig.~\ref{Fig4}a shows the region
$43\fdg3 \times43\fdg3$; the density step is chosen in such a way
that the areas external to the cluster appear better and
the cluster's corona becomes noticeable. This map
was built using the KDE 2D method~\citep{S8} with the parameter
$h=60\arcmin$. The density fluctuations on this map are mainly
due to the field stars. Fig.~\ref{Fig4}b shows the region
$10\degr\times10\degr$, this is the cluster core and the inner part
of its corona. When constructing this map, the parameter
$h=20\arcmin$ was used. It can be seen that the corona
of the cluster is elongated approximately from {southeast}
to {northwest} at an angle close to 45$\degr$. When passing
to denser parts of the corona (Fig.~\ref{Fig4}b), the general
direction of elongation is preserved, but the angle becomes
somewhat steeper, closer to 60$\degr$. The direction of elongation
is in good agreement with the average direction of motion
of the cluster, which is shown by the arrow~\citep{Lod}.
The density level gradation on this map is chosen to show
the contribution of the cluster stars and weaken
the influence of the field stars.

To determine the average density of the number of field stars
and the radius of the cluster, the surface density profiles
were constructed following the technique described in~\citet{S6}.
Fig.~\ref{Fig5} shows surface density profiles obtained with
the parameter $h=80\arcmin$. The choice of the value of $h$
was also made by the method described in~\citet{S6}. In Fig.~\ref{Fig5}a
the density profile of the cluster is given in the axes
``decimal logarithm of distance from the center---decimal
logarithm of density''. Fig.~\ref{Fig5}b illustrates
the determination of the cluster radius and the average density
of the number of background stars:
$R_{\rm cl}=10\fdg9\pm0\fdg3$ $(26.3\pm0.7$\,pc)
and $\overline F_b=0.000562\pm0.000037$ stars per
square arc minute, respectively. Only the region near
the cluster boundary is shown. The thick solid line is
the density profile, and the thick dashed lines are
the $\pm1\sigma$ confidence interval. The thin dashed line shows
the density profile plotted at $h=20\arcmin$. It demonstrates
the correctness of the choice of the parameter $h=80\arcmin$,
since the density profile at this value of the parameter
follows the average trend of the profile plotted at
a much lower half-width. In this case, a change in the parameter
$h$ does not lead to a noticeable change in the estimate
of the cluster radius (the line corresponding to
the profile with $h=20\arcmin$ intersects the line
of average background density at a distance from the cluster radius
that is less than the error in determining the radius).
We also estimated the radius of the cluster core
as the outer radius of the zone of the maximum in absolute value
gradient of the surface density profile: {$R_{c}=2\fdg6$ $(6.2$\,pc).}

After that, the luminosity function (LF) of the cluster
was obtained by the statistical method described in~\citet{S3,S8,S9}.
A ring with an inner radius of $10\fdg9$ and an area
equal to the area of a circle with a radius of $10\fdg9$ was taken
as the comparison area. The one-dimensional KDE method
with the half-width parameter $h=1^{\rm m}$ was used.
Luminosity functions for the cluster core and corona
were obtained in a similar way. The cluster luminosity function
is shown in Fig.~\ref{Fig6}a.

We used the luminosity function and the {mass-luminosity} ratio
from the isochrone tables~\citep{S10} to determine
the mass function (MF) according to the method described
in~\citet{S9}. The MF was obtained for a unit mass interval
of a star and per unit volume. The MF for the cluster
on a logarithmic scale is shown in Fig.~\ref{Fig6}b.
Fig.~\ref{Fig6}$c$ shows the mass functions for the cluster core
and corona, normalized to unity. It is seen that
the relative abundance of stars with $m>1M_{\odot}$
in the corona is lower in comparison with the cluster core.
In turn, the relative abundance of stars with $m<0.4M_{\odot}$
in the cluster's corona is higher than in its core.

Integration of the mass function gives the number of stars
in the cluster $N_{\rm cl}=1542\pm121$ and its mass
\mbox{$M_{\rm cl}=855\pm104\; M_{\odot}$.}
For the cluster core \mbox{$N_{c}=1097\pm77$,}
\mbox{$M_{c}=665\pm71\; M_{\odot}$}.
The {mass--luminosity} dependence by~\citet{S10} can be used
only for stars with magnitudes $m_G\geqslant4.0^{\rm m}$,
therefore, for the 6 brightest stars in the Pleiades,
the mass estimate was obtained through average mass
of the star \mbox{$\overline m=4.66 M_{\odot}$,} also determined
from the isochron tables~\citep{S10}. The obtained estimate
of the number of stars in the cluster can be considered
as the total number of Pleiades stars within the radius
of the cluster corona $R_{\rm cl}=10\fdg9$ up
to the limiting value $m_G=18^{\rm m}$. Incompleteness
can only be associated with the incompleteness of the Gaia data
or with large errors in the parameters of the stars,
as a result of which the star could not be included in our sample.

We estimated the slope of the MF. For stars with mass $m>1M_{\odot}$
it turned out to be equal to {$-2.89\pm$0.03,} for stars
with mass {$m\in[0.5;1]M_{\odot}$}---{$-2.20\pm 0.04$.}
In the region of bright stars, the slope is much greater
than that of the initial Salpeter MF ($-2.35$), but, nevertheless,
both values are quite consistent with the initial
mass function of~\citet{Kroupa}. It should be noted that a large slope
of the Pleiades MF for bright stars was also obtained earlier:
$-2.74\pm0.07$~\citep{Taff} and {$-2.71\pm
0.27$}~\citep{vanLeeuwen}.

\begin{figure*}
   \centering
 \includegraphics[scale=0.31]{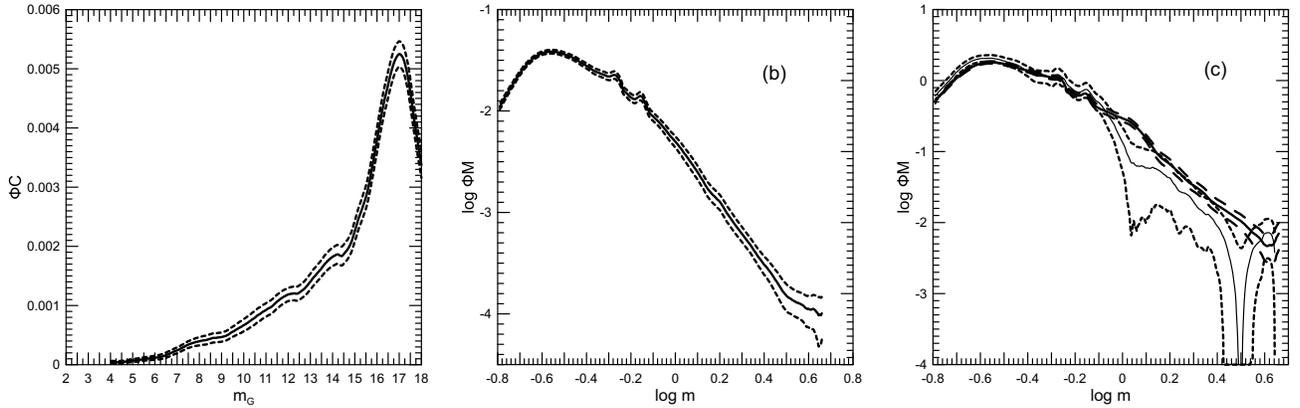}
 \caption{ Luminosity function (a) and mass function (b) of the Pleiades cluster.
 The mass of the star $m$ is given in $M_{\odot}$, the values of the
 LF---in st.mag$^{-1}$~pc$^{-3}$, values of MF---in
 $M_{\odot}^{-1}$pc$^{-3}$. Panel (c) shows the mass function
 for the cluster core (thick lines) and for the corona (thin lines),
 normalized to 1. The dashed lines show the confidence intervals
 \mbox{$\pm1\sigma$ wide.}}
 \label{Fig6}
\end{figure*}

To study the three-dimensional structure and kinematics of
the Pleiades cluster, on the basis of an initial sample
of 47\,195 stars, a sample of stars of probable cluster members
was selected. The procedure was as follows. We narrowed
the intervals for parallaxes and proper motions,
controlling the appearance of the color--magnitude
diagram (CMD). At the same time, as few stars of the field
should remain on the CMD as possible, while retaining
the largest possible number of cluster stars (that is,
stars on the general sequences on the diagram).
The following restrictions were finally adopted
on the values of trigonometric parallaxes and the components
of proper motion of stars: {$\pi_{r}\in[4.6,10]$\,mas,}
proper motion in right ascension
\mbox{$\mu_{\alpha}\in[14,26]$\,mas\,year$^{-1}$,}
proper motion in declination
\mbox{$\mu_{\delta}\in[-51,-40]$\,mas\,year$^{-1}$.}
The width of the intervals was chosen in such a way
that the six brightest stars of the Pleiades got there,
the parallaxes and components of proper motion of which
significantly differ from the average values (see above).
The stars were selected within a circle with a radius
\mbox{$R_{\rm cl}=10\fdg9$} at \mbox{$m_G\leqslant18^{\rm m}$.}
In addition, some stars were excluded from the sample based on their
positions on the CMD
below the main sequence of the cluster.
The resulting sample contains 1391 stars. This number is less
than the statistical estimate of the total number of stars
in the Pleiades (see above). Probably, due to large errors,
some of the stars---cluster members have parallaxes
and proper motions outside the indicated intervals.

In order to estimate the cluster membership probability of
the selected stars, the same selection criteria
were applied to the entire original sample of stars.
As a result, 1965 stars were selected. This means that
there are {$1965-1391=574$} stars outside the cluster's circle.
An area of the figure, bounded by the initial values
of the equatorial coordinates, was determined
by the {Monte Carlo} method with uniformly throwing points
on the sphere (we were throwing one million points) and
turned out to be equal to 0.47152 steradians.
The area occupied by the cluster was defined
as the surface area of a sphere segment,
equal to 0.11336 steradians. Thus, the stars outside
the cluster circle are located on an area of 0.35816 steradians,
which gives an average density of field stars equal to
{$F_{\rm bg}=$1602.64} stars per steradian. Now we can estimate
the number of field stars inside the circle of the cluster
with the used selection criteria, it is equal to 182.
Our estimate of the cluster membership probability
for stars selected within the circle of the cluster is:
\mbox{$P_m=(1391-182)/1391\approx0.87$.}

Thus, our sample of likely cluster members contains approximately
$13$\% of field stars. On the other hand, this sample did not include
approximately $10$\% of the 1542 members of the cluster (see above).
The relative abundance of the field stars changes with distance
from the center of the cluster. In the core, it is much less
than in the corona. It is interesting to compare our sample
with the sample of~\citet{Lod}, also obtained from Gaia\,DR2 data.
This sample \mbox{contains 1412~stars} within a circle of radius
\mbox{$R_{\rm cl}=10\fdg9$} at \mbox{$m_G\leqslant18^{\rm m}$.}
Analysis of the samples showed that they contain 1243~common stars.
Consequently, the samples coincide by approximately $90$\%.
The difference is mainly caused by different selection criteria
and different values of the coordinates of the cluster center.
Nevertheless, our sample can be considered as substantially
complete ($10$\% of ``lost stars'') and substantially
``clean''\,($13$\% of field stars). Further, we will use subsamples
from this sample, selecting stars in a smaller volume of space
and with more accurate determinations of velocities and distances
to stars in accordance with the first approach (see above).
This will always lead to a reduction in the absolute and relative
number of field stars.

\section{DATA ON STARS---PROBABLE MEMBERS OF THE OSC PLEIADES}
\label{R4} In order to study the cluster kinematics, we selected
the subsamples from the sample containing 1391~stars
in the vicinity of the Pleiades cluster and described in the previous section. Firstly, we selected {$N_I=565$}
members of the cluster (with magnitudes {$m_G\le 16\fm03$}
at distances {$d\degr\le 2\fdg5$} from the center of the cluster).
In this sample, the errors {$e_{v_{t}}=\sqrt{e_{v_x}^2+e_{v_y}^2}$}
of velocities \mbox{${v_{t}}=\sqrt{{v_x}^2+{v_y}^2}$}
do not exceed {2.36\,km\,s$^{-1}$,} which for the Pleiades cluster
corresponds to {3.65\,mas\,year$^{-1}$.} Secondly,
we selected {$N_{II}=395$} stars with errors \mbox{$e_{v_{t}}\le
0.177$\,km\,s$^{-1}$} from the $N_I$ sample. These turned out to be stars with
\mbox{$m_G\le15^{\rm m}$} and errors \mbox{$e_r\le 1.5$\,pc}
in the distances $r$. Average errors are
\mbox{$\overline{e_r}\simeq 1.7\pm 0.1$\,pc}
in the sample {$N_I=565$} stars and
\mbox{$\overline{e_r}\simeq 1.00\pm 0.01$\,pc}
in the sample {$N_{II}=395$} stars.
In the sample {$N_I=565$} {member stars} of the cluster,
according to Gaia\,DR2 data, we have selected 74 stars
with radial velocities $V_r$, the errors of which are
\mbox{$e_{V_r}\le 1.0$\,km\,s$^{-1}$.} Average radial velocity
of the stars in the cluster core for these 74 objects is
\mbox{$\overline{V_r}=5.86\pm 0.13$}~km\,s$^{-1}$;
after the correction of $V_r$ for the effect of the transverse motion
of the cluster turned out to be
\mbox{$\overline{V'_r}=5.77\pm 0.13$}~km\,s$^{-1}$.

Thirdly, from the sample of 1391 {member stars} of the cluster with magnitudes
{$m_G\le 18^{\rm m}$} at distances \mbox{$d\degr\le 10\fdg9$} from
the cluster center, for which $e_{v_{t}}\le 2.36$~km\,s$^{-1}$ and
{$e_r\le 38.9$\,pc,} we selected {$N_{III}=550$} stars with {$m_G\le
17^{\rm m}$, $e_{v_{t}}\le 0.177$\,km\,s$^{-1}$,} $e_r\le
1.51$\,pc, $\overline{e_r}\simeq 1.00\pm 0.01$\,pc.
In the sample \mbox{$N_{III}=550$} {member stars} of the cluster,
according to Gaia\,DR2 data, we have selected 97~stars with
radial velocities, the errors of which are
\mbox{$e_{V_r}\le 1.0$\,km\,s$^{-1}$.}\,Average radial velocity
 of cluster stars for\,these 97\,objects\,$\overline{V_r}$=$5.66\pm
0.20$\,km\,s$^{-1}$; after correcting $V_r$ for the effect
of transverse motion of the cluster---\mbox{$\overline{V'_r}=5.37\pm 0.13$\,km\,s$^{-1}$.}
Thus, the correction of \mbox{$\overline{V_r}$} for the effect
of the transverse motion of the cluster changes the value of $\overline{V_r}$
only within the error, which is due
to the symmetry of the cluster and the corrections $\Delta V_r$
of the radial velocities of the stars about the axis $x'$
perpendicular to the direction of motion of the cluster
in projection onto the sky plane, see Fig.~\ref{Fig1}
(74 stars from $N_{I}$ with {$e_{V_r}\le 1.0$\,km\,s$^{-1}$}
are located in the sky plane more symmetrically
about the axis $x'$ than 97 stars from $N_{III}$).
The values \mbox{$\overline{\Delta
V_r}=\overline{V_r(I)}-\overline{V_r(III)}=0.20\pm0.24$\,km\,s$^{-1}$}
and \mbox{$\overline{\Delta
V'_r}=\overline{V'_r(I)}-\overline{V'_r(III)}=0.40\pm0.18$\,km\,s$^{-1}$.}
The indices of the star samples $N_{I}$ and $N_{III}$ used
in the calculation of $\overline{V_r}$ and $\overline{V'_r}$
are indicated in brackets.
Probably, the field stars in the considered samples
only insignificantly affect the estimates of the values
$\overline{\Delta V_r}$ and $\overline{\Delta V'_r}$
because field star contamination in the $I$ and $III$ samples
are less than 1\% and less than 8\%, respectively (see below).

The equality of the values $\overline{\Delta V'_r}$ and $\overline{\Delta
V_r}$ indicates the symmetry in the arrangement of the stars
under consideration relative to the $x'$ axis and the relative motion
along the line of sight of the selected groups of stars
in the samples $N_{I}$ and $N_{III}$ with the velocity
{$\overline{\Delta V'_r}\simeq\overline{\Delta V_r}$.}
Since in our case \mbox{$\overline{\Delta
V'_r}-\overline{\Delta V_r}=0.2\pm0.3$~km\,s$^{-1}$,}
it seems reasonable to consider the relative radial velocity
of these two groups of stars to be equal: $(\overline{\Delta
V'_r}+\overline{\Delta V_r})/2=0.30\pm0.15$~km\,s$^{-1}$
(the cluster core is moving away from the Sun faster
than the cluster corona).
The average radial velocity of the entire cluster is \mbox{$V_{\rm
r,cl}=5.67\pm0.08$~km\,s$^{-1}$.}
In the work of \citet{Lod}, this value for practically complete sample
of stars-members of the Pleiades cluster is obtained equal to
{$V_{\rm r,cl}=5.67\pm2.93$~km\,s$^{-1}$.}
Probably, the large error of $V_{\rm r,cl}$ in~\citet{Lod} is caused
by large errors in the radial velocities $V_r$ of the stars
under consideration (since the sample is complete) and
the neglect of the contribution to $V_r$ of the transverse motion
of the cluster in the sky plane.
Note that the gravitational potential and the equations of motion
of stars in the OSC are nonlinear, and the dynamic evolution
of the OSC is determined by various types of instabilities.
Therefore, the use of the most accurate data on the coordinates
and velocities of stars is the only possible way to obtain
correct results on the dynamic mechanisms operating in the OSC
(for example, inaccuracies in determining the parameters of just
one dynamically active pair of stars in the OSC can affect
the predicted evolution of the entire cluster, consisting
of several hundred stars; instead of the cluster expansion,
one can get a forecast of its contraction~\citep {D}).

The 97 stars with {$e_{V_r}\le 1.0$}~km\,s$^{-1}$ considered here
from the $N_{III}$ sample are located in the sky plane
at distances {$d\degr\le 8\fdg56$} from the cluster center.

We accepted the estimate of the distance to the cluster
{$r_{\rm cl}=136.4\pm 0.2$~pc} (see above), the distance modulus
$(m_V-M_V)_0=(m_G-M_G)_0=5\fm84\pm 0\fm16$}~\citep{M24},
complete absorption {$A_G\simeq 0\fm08$} ~\citep{M24}.
The relationship between $A_G$ and the total absorption $A_V$
in the $V$ band was taken from
{\url{http://stev.oapd.inaf.it/cgi-bin/cmd_3.3}},
where it was defined on basis of the work of~\citet{M25}.
In this case, the average masses of stars in the considered samples
were obtained equal to
\mbox{$\overline{m}_I=0.88\pm0.17~M_{\odot}$,}
\mbox{$\overline{m}_{II}=1.04\pm0.21~M_{\odot}$,}
\mbox{$\overline{m}_{III}=0.82\pm0.11~M_{\odot}$,} which
corresponds to the total masses of the selected stars
$M_I=497\pm96~M_{\odot}$, $M_{II}=411\pm83~M_{\odot}$,
$M_{III}=451\pm61~M_{\odot}$, respectively.
The quantities $\overline{m}_i$ and $M_i (i=I,II,III$)
were used to calculate the cluster parameters included in
formula (\ref{eq3}) and in the coefficients of equation (\ref{eq4}).

When studying the stars of the sample $II$, we noticed
a linear dependence {$m_G=(11.19\pm0.07)^{\rm m}+(0.22\pm0.04)^{\rm
m}r_s$/pc,} where the distance of a star from the center of the cluster
is {$r_s\le 6.9$~pc,} which indicates the segregation
of the most massive stars in the Pleiades by $r_s$
in three-dimensional space and the decrease in the masses $m$
of such stars with distance from the center of the cluster
(the sample $II$ contamination with field stars is less than 1\%, see below).

\begin{figure*}
   \centering
\includegraphics[scale=0.55]{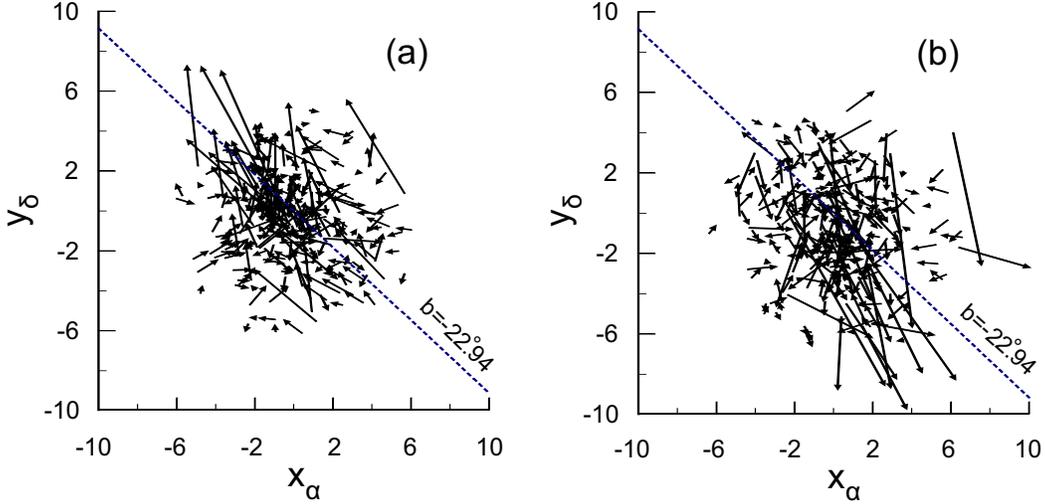}
\caption{The projections of the velocity vectors of stars in
the Pleiades cluster core onto the sky plane
$(x_{\alpha},y_{\delta})$, multiplied by 1~Myr
(given in parsecs, 1\,pc~Myr$^{-1}\simeq 1$~km~s$^{-1}$);
(a) for stars with $r<r_{\rm cl}$, (b) for stars with
$r>r_{\rm cl}$; $r$ and $r_{\rm cl}$ are the distance of stars
and the cluster from the Sun in~pc, respectively.
The line of galactic latitude $b=-22\fdg94$ is shown by dashed line.}
 \label{Fig7}
\end{figure*}

\section{ESTIMATES OF THE ROTATION PARAMETERS OF THE OSC PLEIADES}
\label{R5} Let $x_{\alpha}$ and $y_{\alpha}$ be the axes
of the tangential coordinate system centered at the center
of the cluster~\citep{S12}. The projections of the velocity vectors
of the stars in the Pleiades cluster core onto the plane
$(x_{\alpha},y_{\delta})$ are obtained from the data on the coordinates
$(\alpha,\delta)$ and the proper motions $(\mu_{\alpha},\mu_{\delta})$
of stars of the sample $I$ (see Section~\ref{R4});
the radius of the area occupied by sample stars relative to the center
of the cluster on the celestial sphere is {$d\degr=2\fdg5$;}
calculations of $P_m$, similar to those performed in Section~~\ref{R3},
for a sample of Pleiades stars with $d\degr\le 2\fdg5$ result
in probability \mbox{$P_m\simeq 0.99$.}
Thus, the sample $I$ considered here is practically not contaminated
with field stars. The total number of cluster stars in an area
with a radius of $2\fdg5$ according to the method from Section~\ref{R3}
is obtained equal to 958, the completeness of the sample $I$
is \mbox{$\Pi_I\simeq 565/958\simeq 0.59$;} for the sample $II$---\mbox{$\Pi_{II}\simeq 395/958\simeq 0.41$;} in the case of
the sample $III$, the value $P_m=0.92$, the number of {member stars}
of the cluster in this sample ($d\degr\le 8\fdg56$) is 1347,
therefore \mbox{$\Pi_{III}\simeq 550/1347\simeq 0.41$.}

The average motion of stars with $r<r_{\rm cl}$ and $r>r_{\rm cl}$
in Fig.~\ref{Fig7} $a,b$ is directed on the plane $(x_{\alpha},y_{\delta})$
in opposite directions, the rotation of the core of the Pleiades cluster
is ``prograde'' (directed in the same direction as the rotation
of the Galaxy). Passing to the coordinate system $(x,y)$
(see explanations before formula~(\ref{eq6})), we calculate
the values \mbox{$\Delta{\overline{v}}_x$,~$\Delta{\overline{v}}_y$}
and find for the cluster core
\mbox{$|\Delta{\overline{v}}_x|=0.53\pm0.07$~km~s$^{-1}$,}
\mbox{$|\Delta{\overline{v}}_y|=0.18\pm0.04$~km~s$^{-1}$} and
the value {$v_c=\overline{v_{x,y}}=0.56\pm0.07$\,km\,s$^{-1}$}~ at~
the~~ average~~ distance \mbox{$\overline{d}=2.95\pm0.07$~pc}
from the cluster center in the sky plane (when determining the sign of
the values $\overline{v_{x,y}}$ and $\omega_z$, we used data
on the motion of stars in the cluster core relative to the $z$ axis
in the $(x,y)$ plane and screw-gimlet rule, see~\citet{A20} p.~31).
The angle between the positive direction of the $y$ axis
and the projection of the axis of rotation of the cluster core
onto the $(x,y)$ plane is equal to
$\varphi=\arctan(|\Delta{\overline{v}}_y/\Delta{\overline{v}}_x|)=18\fdg8\pm4\fdg4$.
The value $v_c$ can be called the equatorial rotation velocity
of the cluster core (the corresponding vector ${\bf v}_c$ is located
in a plane perpendicular to the rotation axis of the cluster core;
this can be verified by rotating the axes $x$, $y$ by an angle $\varphi$
relative to the $z$ axis, leading to a system of axes $x''$, $y''$, $z''$,
in which the vector ${\bf v}_c$ is parallel to the $x''$ axis
(or antiparallel to it, see Fig.~\ref{Fig7}a,b), and the next rotation
of the $y''$, $z''$ axes by the angle $\vartheta$ relative
to the $x''$ axis results in the coordinate system $x'''$,
$y'''$, $z'''$, in which the plane $y'''=0$ is equatorial).

According to formula (\ref{eq6}), the angle of inclination
of the axis of rotation of the core of the Pleiades OSC
to the sky plane $\vartheta$ is $43\fdg2\pm4\fdg9$.
Used in (\ref{eq6}) to estimate $\vartheta$, the values
\mbox{$|\Delta{\overline{v}}_z(x)|=0.206\pm0.119$\,km\,s$^{-1}$,}
\mbox{$|\Delta{\overline{v}}_z(y)|=0.011\pm0.126$\,km\,s$^{-1}$}
have been obtained from the data on the radial velocities $V_r$
(corrected for the motion of the OSC in the sky plane) with errors
\mbox{$e_{V_r}\le 1.0$\,km\,s$^{-1}$} for 74~stars
from the sample $I$. Note that the average ${|z|}$ for these stars is
$\overline{|z|}\simeq 3.13\pm0.40$\,pc $\simeq\overline{d}$ (see above).
The cluster core in the sky plane has a shape close to spherical one
(Fig.~\ref{Fig7}). Therefore, in formula (\ref{eq6}),
we replaced the angular velocities of the stars by circular ones.
In this case \mbox{$\overline{{ v}_{x,z}}=0.57\pm0.08$\,km\,s$^{-1}$}
and \mbox{$\overline{{ v}_{y,z}}=0.18\pm0.04$\,km\,s$^{-1}$.}

Linear regression dependencies on $x$ and $y$ of the radial velocities
of stars $V_r'$, corrected for the motion of the OSC core
in the sky plane, for \mbox{$x=y=5.5$\,pc} lead to the values
$V_r'(x)-V_r'(x=0)=-0.127\pm0.300$\,km\,s$^{-1}$ and
$V_r'(y)-V_r'(y=0)=0.097\pm0.307$ km\,s$^{-1}$,
respectively. Then the equatorial velocity is
$v_c\simeq 0.16\pm0.30$\,km\,s$^{-1}$.
Thus, at distances from the center of the cluster of the order of
\mbox{$d=5.5$}\,pc, the core rotation velocity is close to zero
(some of the core stars participate in the ``prograde'',
and some---in the ``retrograde''\, motion about the axis of rotation).
Similar calculations for 97 stars with \mbox{$e_{V_r}\le
1.0$\,km\,s$^{-1}$} from the sample~$III$ of the cluster
lead to the velocity of the ``retrograde''\, rotation of the cluster
\mbox{$v_c\simeq 0.48\pm0.20$\,km\,s$^{-1}$}
at the distance \mbox{$d=7.1$\,pc} from its center. The angle
between the projection of the cluster's rotation axis onto
the sky plane and the direction of increasing galactic latitude $b$ is
{$\varphi=\arctan[|V_r'(y)-V_r'(y=0)|/|V_r'(x)-V_r'(x=0)|]=37\fdg8\pm26\fdg4$.}
Thus, with an increase in the distance from the center of
the Pleiades cluster for \mbox{$d>5.5$\,pc}, the rotation
of the cluster becomes ``retrograde'', which indicates
the possible stability of the trajectories of motion of some stars
on the cluster periphery in the joint field of forces
of the cluster and the Galaxy. Note that in the halo of OSC
models~\citep{A14}, see Fig.~1 from~\citet{A14}, the ``retrograde''\,
motions of the cluster stars also dominate.

\section{ESTIMATES OF THE VELOCITY DISPERSIONS OF STARS IN THE PLEIADES}
\label{R6}

\begin{figure*}
   \centering
\includegraphics[scale=0.63]{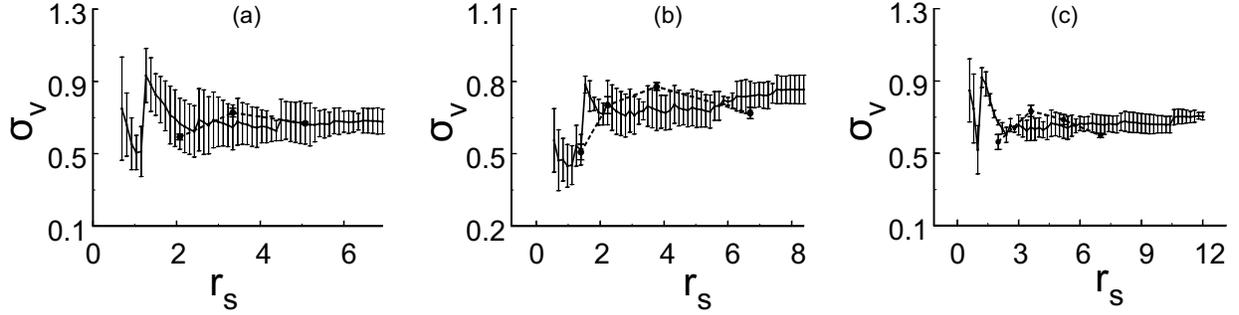}
\caption{Dependences of the rms velocities of stars $\sigma_v$ on $r_s$
inside a sphere of radius $r_s$ centered at the center of
the Pleiades cluster: (a) for the core stars with
$m_G<15^{\rm m}$ (the sample $II$), (b) for the core stars with
$m_G<16^{\rm m}$ (the sample $I$), (c) for the stars with
$m_G<17^{\rm m}$ (the sample $III$). Dashed lines correspond
to the  $\sigma_v=\sigma_{v,J}$, obtained from
condition~(\ref{eq2}) of gravitational instability of the central
regions of the cluster located within the sphere of radius $r_s$.
The solid lines show the radial dependences of the $\sigma_v$
 determined from the data on the proper motions of stars.}
 \label{Fig8}
\end{figure*}

\begin{figure*}
   \centering
\includegraphics[scale=0.67]{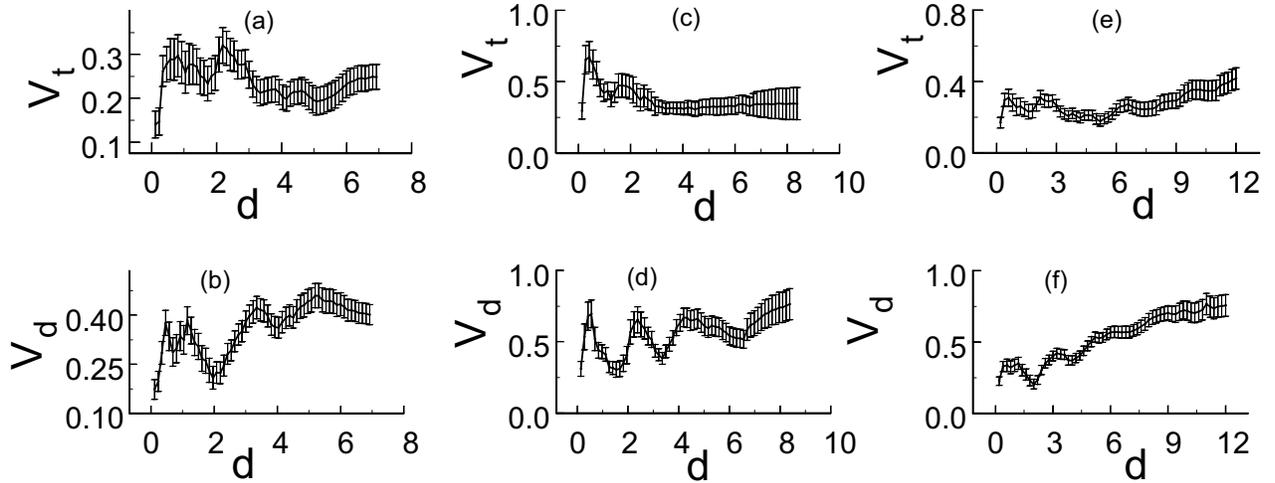}
\caption{Dependences of the moduli of the tangential $V_t$ and radial
{$V_d$-components} of velocities of the stars of the Pleiades cluster
in the sky plane on the distance $d$ to its center;
$a,d$ are obtained from the sample $II$, $b,e$---from the sample $I$,
$c,f$---from the sample $III$ of the cluster stars.}
 \label{Fig9}
\end{figure*}

When constructing the dependences of the root-mean-square
velocities (Fig.~\ref{Fig8}), the values $R_m=6.9$\,pc, 8.4\,pc,
12.0\,pc  were used (see Section~\ref{R2} and data on the parameters
of stars in the samples $II$, $I$, $III$, respectively). The condition
of gravitational instability \mbox{$\sigma_v\le\sigma_{v,J}$}
in the cluster is satisfied in the regions with \mbox{$r_s\in [2.2,5.2]$\,pc}
in Fig.~\ref{Fig8}a and $r_s\in [2.2,5.7]$\,pc in
Fig.~\ref{Fig8}$b,c$.
In these ranges of $r_s$,  approximately $39.4$\% of the stars
in the sample $II$ (the brightest and most massive stars of the cluster),
approximately $60.5$\% of the stars in the sample $I$, and about $52.6\%$
of the stars in the sample~$III$ are contained.
Near the center of the cluster, the values of $\sigma_v$ change
noticeably with increasing $r_s$, and it is impossible to obtain
data on the gravitational instability due to the highly irregular
structure of the cluster, large deviations of the cluster
from the equilibrium state, and large errors in the velocities
of the stars in this region. At \mbox{$r_s< 2$\,pc}, the core
of the Pleiades cluster by the time of observation is formed
from approximately spherical layers with a common center and
different values of $\sigma_v$ of stars in the layers adjacent to $r_s$.
It is possible that a wave process of transfer of kinetic energy
of stellar motions in the radial direction takes place in the cluster.
In OSC models~\citep{A21} such a process could not be noticed,
since in this article \citetext{\citealp*[Fig.~1b,c, Tables~1b,c]{A21}}
only the average values of $\sigma_v^2$ over
the oscillation period of the regular field of the cluster were considered.
The wave process of the transfer of the kinetic energy of stars
in the radial direction in a non-stationary cluster is described
in more detail in~\citet{Ax} within the framework of the collisionless
``water bag'' model using numerical integration of the Vlasov system
of equations for an isolated {spherically symmetric} cluster of stars.

According to Fig.~\ref{Fig8}, for $r_s>2$\,pc the values of $\sigma_v$
in the considered samples of stars, on average, increase with increasing
$r_s$, which indicates the nonstationarity of the cluster in the field
of regular forces. The dependences of the modules $V _t$ and $V_{d}$ on $d$
in Fig.~\ref{Fig9} contain a number of periodic oscillations
that go beyond the errors of $V_t$ and $V_{d}$ by an amplitude,
which also indicates the nonstationarity of the cluster
in the field of regular forces.

In the process of evolution of non-isolated and nonstationary (in a regular field) models
in a regular field of OSCs at $t>(0.3$--$0.6)\tau_{v.r.}$,
entropy is produced not in the core, but on the cluster periphery
(see Fig.~1 from \cite{A2}) due to numerous weak stellar encounters.
In the cores of nonstationary OSCs, evolution is determined
by the interactions of stars with an alternating force field
of the cluster, the action of gravitational instability, which
leads to an increase in the correlation of stellar motions~\citep{A2}.
Stars with the highest energies that have received additional energy
during violent relaxation in the core enter the periphery
of the cluster, where, under the influence of approaches
with stars of a rather densely populated halo, they acquire
additional angular momentum and lose the opportunity to return
to the cluster core (in contrast to quasi-stationary and isolated
systems close to virial equilibrium, in which entropy is produced
mainly due to close stellar encounters in the cluster core).
As a result, stars with the highest energies leave the core,
which leads to the cooling of the core of the nonstationary cluster.

\begin{figure*}
   \centering
\includegraphics[scale=0.47]{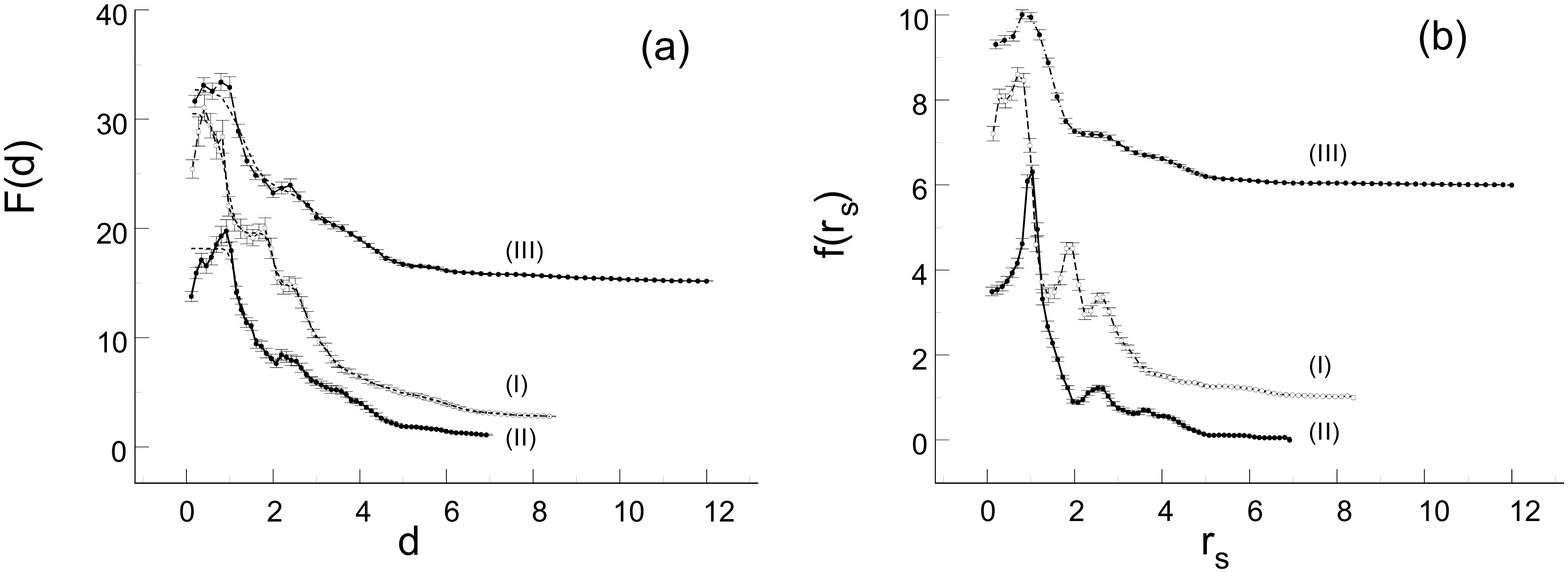}
\caption{(a) Dependences of the apparent density $F(d)$ of the number of stars
in the Pleiades cluster on the distance $d$ to its center;
the dotted lines show $F(d)$ smoothed by the method from~\citep{M16},
used to obtain the spatial density distributions $f(r)$
in the cluster, see Section~\ref{R2}.
Labels $I$, $II$, $III$ in Fig. (a) and (b) denote
sample numbers. The $I$ curve is shifted up along the ordinate axis
by 2~pc$^{-2}$, the $III$ curve by 15~pc$^{-2}$.
(b) Dependences of the spatial density $f(r_s)$ of the number
of stars in the Pleiades cluster on the distance $r_s$ to its center,
obtained according to the method described in Section~\ref{R2}.
The curve $I$ is shifted up along the ordinate axis by 1 pc$^{-3}$,
the curve $III$---by 6~pc$^{-3}$.}
 \label{Fig10}
\end{figure*}

According to Fig.~\ref{Fig10}, the brightest and most massive
cluster stars from the samples $II$ and $I$ near the cluster center
are distributed most unevenly, the centers of their distributions
do not coincide, local maxima of the apparent density
in the sky plane around the cluster center are achieved
near some closed curves, which indicates a deviation from
the equilibrium state and nonstationarity of the cluster
in the field of regular forces. Equatorial coordinates
of the center of distribution of stars in the sample $II$:
\mbox{$\alpha_c^{II}=56\fdg240\pm0\fdg003$,}
\mbox{$\delta_c^{II}=24\fdg420\pm0\fdg003$,} for stars of the sample
$I$: \mbox{$\alpha_c^{I}=56\fdg708\pm0\fdg003$,}
\mbox{$\delta_c^{I}=24\fdg420\pm0\fdg003$,} for stars of the sample
$III$: $\alpha_c^{III}~=~56~\fdg208~\pm~0\fdg003$,
\mbox{$\delta_c^{III}=24\fdg418\pm0\fdg003$,}
respectively (when determining the values of $\alpha_c$, $\delta_c$,
we used the method described in Section~\ref{R3} of this article).
Note that the difference in the coordinates of the center
of distribution of stars by $0\fdg5$ in the case of the Pleiades
corresponds to a distance of about $1.19$\,pc from the center
of the cluster in the sky plane. The depression of the density
$f$ in the center of the M\,37 cluster, which is very close
in appearance to that shown in Fig.~\ref{Fig10}b for
the sample $II$, was noted in the book of~\citet[p.~317]{M17}.

\begin{table*}
   \centering
 \caption{Estimates of the structural and dynamic parameters of the Pleiades cluster.
  The columns indicate:
 (1) $m_G$ stellar magnitude according to Gaia;
 (2) dynamic velocity dispersion by formula (3);
 (3) $R_u=\langle 1/r_s \rangle^{-1}$, $r_s$---distance of a star from
     the center of mass of a cluster with averaging over all stars of the cluster;
(4) average cluster radius (the mean distance between two cluster stars with averaging over all pairs of stars in the cluster);
(5) average square distance of a star from the center of the cluster;
 (6) $R_m$---radius of the region in which the velocity field of cluster stars
    is studied;
  (7) dynamic mass;
 (8) tidal radius}
 \label{Tab 1}
 \medskip
\begin{tabular}{c|c|c|c|c|c|c|c}
\hline $m_G$ &  {$\sigma_{v,d},$~km~s$^{-1}$} &  {$R_u,$~pc} & {$\overline{R},$~pc} & {$\overline{r_s^2},$~pc$^2$} & {$R_m,$~pc} & {$M_d,~M_{\odot}$} & {$R_t,$~pc}\\
\hline (1) & (2) & (3) & (4) & (5) & (6) & (7) & (8)\\
\hline
$m_G\le15^{\rm m} $&   $0.80\pm0.11$ & $2.53\pm 0.02$ & $2.72\pm 0.04$  & $15.1\pm 0.4$ &   $6.9$ & $441\pm 123$ &$10.5\pm 0.3$\\
$m_G\le16^{\rm m} $ &  $0.71\pm 0.07$ & $2.73\pm 0.03$ & $2.93\pm 0.03$  & $18.52\pm 0.43$  &  $8.4$ & $371\pm 80$ & $9.9\pm 0.2$ \\
$m_G\le17^{\rm m} $&  $0.72\pm 0.02$ & $3.25\pm 0.02$ & $3.54\pm 0.04$  & $32.92\pm 1.07$ &  $12.0$ & $462\pm 29$ & $10.6\pm 0.1$\\
\cline{2-8}  &  $ - $ & $3.59\pm 0.02$ & $4.02\pm 0.06$  & $54.48\pm 2.31$ &  $20.5$ & $513\pm 33$ & $11.0\pm 0.1$\\

\hline
\end{tabular}
\end{table*}

Density waves in Fig.~\ref{Fig10}, which exceed by an oscillation amplitude the error limits
of the values $F(d)$ and $f(r)$,
also indicate the nonstationarity of the Pleiades cluster
in the field of regular forces.

The radial distributions of the spatial density of the number of stars
$f(r_s)$ and the mass \mbox{$\rho(r_s)=\overline{m}f(r_s)$}
in the Pleiades were used here to find the values
$f_c=f(r_c)$, $\rho_c=\rho(r_c)$, $M_c$ and
\mbox{$\sigma_v^2=\sigma_{v,J}^2$} by formula (\ref{eq2})
for a range of values $r_s=r_c$ in the vicinity
of the cluster core boundary (see intervals of values of $r_s$
for dashed lines in Fig.~\ref{Fig8}).
We believe that more accurate \mbox{$f=f(r_s)$,}
obtained using the coordinates of stars in the $(x_{\alpha},y_{\delta})$
plane and the assumption of spherical symmetry of the cluster,
are more preferable than the values $f=f(r_s)$, which are obtained
from the data on the three spatial coordinates of the stars,
since the errors $e_r$ in distances to stars are comparable
to the scale of inhomogeneities in $r_s$ and $d$ in
\mbox{$f=f(r_s)$ and $F=F(d)$,}  respectively, see~Section~\ref{R4}
and Fig.~\ref{Fig10}, and when estimating the dynamic state of the OSC,
the force fields arising in the core between stars and groups
of stars are important (see local maxima of the apparent
density near the center of the cluster in Fig.~\ref{Fig4}).

For the Pleiades cluster, the estimates of the structural
and dynamic parameters were acquired according to
data obtained from the considered samples of stars. These parameters are
listed in Table~\ref{Tab 1}.
The methodology from works of~\citet{A8,A18}, briefly described in
Section~\ref{R2} of our article, and formulas (\ref{eq3}) and
(\ref{eq5}) have been used. The value $R_t$
have been obtained from the estimate of the dynamic mass $M_d$
of the cluster for the OSC in a circular orbit in the field of forces
of the Galaxy according to formula (11.13) from \cite{A9}};
the distance of the Pleiades from the center of the Galaxy
is obtained equal to \mbox{$R_G=8643\pm650$\,pc,} the distance
of the Sun from the center of the Galaxy is taken equal to
$R_{\odot}=8500$\,pc~\citep{A8,M32}, the error in calculating
$R_{\odot}$ is taken equal to $0.1R_{\odot}$~\citep{A8},
the values of $\alpha_1$ and $\omega$ were determined according
to the model of the Galaxy potential~\citep{A12}.

Note that with an increase in the limiting values of $m_G$ from
$16^{\rm m}$ to $17^{\rm m}$ and the radius of the investigated region
$R_m$, the sample of stars includes larger number of stars
of the cluster corona, therefore its dynamic mass $M_d$ increases
(probably, the values $M_d$ indicated in Table~\ref{Tab 1}
can be considered as lower estimates of the mass of the cluster
and its corona). For $m_G\le18^{\rm m} $, according to the data
obtained in Section~\ref{R3} of this article, for
\mbox{$M_{\rm cl}=855\pm104 M_{\odot}$} by formula (11.13)
from~\citet{A9}, we find:  \mbox{$R_t=13.01\pm 0.53$\,pc.}
In the works of~\citet{A4,RE1,RE2},
and~\citet{Lod} for the Pleiades, the following estimates
for $R_t$ had obtained: 13.1\,pc, 16\,pc, $16.53\pm1.52$\,pc and 11.6\,pc,
respectively. Note that the estimates of the value of $R_t$
in the papers of~\citet{RE1,RE2,Lod} are based on the approximation
by of the radial profile of the cluster surface density by the King's formula.
Our estimate of $R_t$ is obtained using the cluster mass function.
Determining the values of $R_t$ by different methods
may well give slightly different results.

In~\citet{A22}, corona models were constructed for six numerical
dynamic models of OSCs. According to this work,
the retrograde stellar motions predominate in the coronas; in the interval
of distances from the center of the cluster model $r/R_t\in (1,3]$,
the formation of the density and phase density distributions close
to equilibrium is noted.
The temporal equilibrium of the coronas is due to the balance
of the number of stars entering the corona from the central regions
of the cluster and leaving to the periphery of the corona or beyond.
According to \citetext{\citealp*[Table~3]{A22}}, by the time
$t=3\tau_{v.r.}\simeq1.5\times 10^8$\,years the coronas
of the OSC models contain $(58.2$--$74.8)\%$ of the total number
of stars in the cluster. Up to the distances $r\simeq4R_t$
from the center of the cluster model, the presence of close to periodic
retrograde average motions of a large number of corona stars is noted,
and at the lifetime intervals of the cluster {(91--99)\%} of corona stars
satisfy the criterion of gravitational coupling~\citep{A23}.
The radius $10\fdg9\pm0\fdg3$ ($26.3\pm 0.7$\,pc) and the mass
of the Pleiades cluster, which exceed its tidal radius and dynamic mass,
obtained as a result of star counts in Section~\ref{R3}
of this article, are quite consistent with numerical data
on the parameters of the coronas of the OSC models \citep{A22}.

\section{CONCLUSIONS}
\label{R7}

1. In this paper we evaluate a number of structural, kinematic and dynamic
parameters of the Pleiades OSC.
According to Gaia DR2 data on stars with magnitudes
$m_G\le 18^{\rm m}$ in the sky region $60\degr\times60\degr$
with the center in the center of the cluster, a density, density profile,
luminosity and mass functions of the cluster were constructed.
We obtained: radius of the cluster \mbox{$10\fdg9\pm0\fdg3$,}
radius of its core $2\fdg62$, number and mass of stars in \mbox{the cluster
$1542\pm121$} and \mbox{$855\pm104~M_{\odot}$,} number and mass
of stars in the cluster core \mbox{$1097\pm77$} and \mbox{$665\pm71~M_{\odot}$.}
These parameters characterize the general structure of the Pleiades
cluster and its corona. A complex irregular structure of the cluster
core is noted (see Fig.~\ref{Fig4}), which is more complex
for the subsystem of brighter stars with \mbox{$m_G\le 15^{\rm m}$,}
which indicates large deviations of the cluster core from equilibrium
in the regular field.

2. We mark an additional indication of the non-stationarity of the Pleiades
cluster in the regular field. There are the radial waves of the apparent
and spatial density of the number of stars noted in Fig.~\ref{Fig10},
as well as waves on the dependences on the distance $d$
to the cluster center of tangential and radial components
of the velocity field of the cluster core stars
in the sky plane, marked in Fig.~\ref{Fig9}.

3. Dispersions of the velocities of the stars $\sigma_v$ in the cluster
core increase on average with an increase in the distance $r_s$
from its center (see Fig.~\ref{Fig9}), which is also
a kinematic sign of nonstationarity of the cluster in
the regular field \citep{A21,A2}. The region of gravitational
instability in the Pleiades cluster is located at distances
\mbox{$r_s=2.2$--$5.7$\,pc} from its center and contains
$39.4$--$60.5$\% of the total number of stars in the considered
samples of cluster objects. It is impossible to obtain data
on gravitational instability near the center because of
the highly irregular structure of the cluster, large deviations
from the equilibrium state, and large errors in the velocities
of the stars.

4. We obtained the estimates of the dynamic mass
$M_d\sim(370$--$510)M_{\odot}$
 and tidal radius \mbox{$R_t\sim (10$--$11)$\,pc}
of the Pleiades cluster for different
$m_G<15^{\rm m}$, $16^{\rm m}$, $17^{\rm m}$.
When increasing the limiting magnitudes $m_G$ and the radius $R_m$
of the region under study, the sample of stars includes more and more
stars of the cluster's corona; therefore, the cluster dynamic mass $M_d$
increases. Thus, the radius and mass of the Pleiades cluster,
obtained as a result of star counts in Section~\ref{R3},
exceeding its tidal radius and dynamic mass, are in good agreement
with the numerical data on the parameters of the coronas of
the OSC models~\citep{A22}.

5. The average rotation velocity of the cluster core
\mbox{$v_c=0.56\pm0.07$~km~s$^{-1}$} at distances $d\le4.6$\,pc
from its center was determined by the data on stars with $m_G < 16^{\rm m}$.
The rotation is ``prograde'', the angle between the projection
of the axis of rotation of the cluster core onto the sky plane
and the direction of growth of the galactic latitude b is
\mbox{$\varphi=18\fdg8\pm4\fdg4$,} the angle between the axis
of rotation of the cluster core and the sky plane is
$\vartheta=43\fdg2\pm4\fdg9$. The rotation velocity of the cluster core
at a distance of $d\simeq 5.5$\,pc from its center is close to zero:
$v_c=0.1\pm0.3$~km~s$^{-1}$. According to data on stars with
$m_G<17^{\rm m}$, the velocity of the ``retrograde'' rotation
of the cluster at a distance of $d\simeq 7.1$\,pc from its center
$v_c=0.48\pm0.20$~km~s$^{-1}$, the angle \mbox{$\varphi=37\fdg8\pm26\fdg4$.}
The obtained estimates of the rotation parameters of the Pleiades cluster
can be used in the numerical simulation of the dynamics of OSCs.
The obtained velocities and directions of rotation of the core
and outer regions of the Pleiades cluster are in good agreement
with the data on the rotation of the OSC models \citep{A14,A22}.

\begin{acknowledgments}
The authors are grateful to A.~A.~Popov, a researcher
at the Astronomical Observatory of the Ural Federal University,
who pointed out the possibility of the influence of the OSC motion,
perpendicular to the line of sight, on the radial velocities
of the cluster stars.

This work used data from the European Space Agency (ESA) Gaia mission
({\url{https://www.cosmos.esa.int/gaia}}), processed by the Gaia
Data Processing and Analysis Consortium (DPAC,
{\url{https://www. cosmos.esa.int/web/gaia/dpac/consortium}}).
Funding for DPAC was provided by national institutions, in particular
institutions participating in the Gaia multilateral agreement.

\end{acknowledgments}

\section*{FUNDING}
This work was supported by the
Ministry of Science and Higher Education, FEUZ--2020--0030.  This work
was supported in part by the Act no. 211 of the Government of the
Russian Federation, agreement no. 02.A03.21.0006.

\section*{CONFLICT OF INTEREST}
The authors declare no conflicts of interest.


%

\end{document}